\documentclass[journal]{IEEEtran}

\ifCLASSINFOpdf
\else
   \usepackage[dvips]{graphicx}
\fi
\usepackage{url}

\usepackage{graphicx}
\usepackage{algorithm}
\usepackage{algorithmic}
\usepackage{amsmath,amssymb,amsfonts}
\usepackage{array}
\usepackage{bm}
\usepackage{cite}
\usepackage[varg]{txfonts}
\usepackage{wrapfig}
\usepackage{color}
\usepackage[switch,pagewise,displaymath, mathlines]{lineno}
\newtheorem{definition}{Definition}

\renewcommand{\algorithmicrequire}{\textbf{Input:}}
\renewcommand{\algorithmicensure}{\textbf{Output:}}

\newcommand*{\reviewerA}[1]{{#1}}%
\newcommand*{\reviewerB}[1]{{#1}}%
\newcommand*{\reviewerC}[1]{{#1}}%
\newcommand*{\reviewerCC}[1]{\textcolor{black}{#1}}
\newcommand*{\reviewerD}[1]{\textcolor{black}{#1}}%
\begin{document}

\title{Data-Driven Sensor Selection Method Based on Proximal Optimization for High-Dimensional Data With Correlated Measurement Noise}

\author{Takayuki Nagata, \IEEEmembership{nonmember, IEEE}, Keigo Yamada, \IEEEmembership{nonmember, IEEE}, Taku Nonomura, \IEEEmembership{nonmember, IEEE}, \\ Kumi Nakai, \IEEEmembership{nonmember, IEEE}, Yuji Saito, \IEEEmembership{nonmember, IEEE}, Shunsuke Ono, \IEEEmembership{member, IEEE}
\thanks{T.~Nagata is with Tohoku University, Sendai, 980-8579, Japan (e-mail: nagata@tohoku.ac.jp), 
K.~Yamada is with Tohoku University, Sendai, 980-8579, Japan (e-mail:keigo.yamada.t5@dc.tohoku.ac.jp ),
T.~Nonomura is with Tohoku University, Sendai, 980-8579, Japan (e-mail: nonomura@tohoku.ac.jp),
K.~Nakai is with Tohoku University, Sendai, 980-8579, Japan (e-mail: kumi.nakai@tohoku.ac.jp), 
Y.~Saito is with Tohoku University, Sendai, 980-8579, Japan (e-mail: yuji.saito@tohoku.ac.jp),
S.~Ono is with the Department of Computer Science, Tokyo Institute of Technology, Tokyo, JAPAN (ono@c.titech.ac.jp).}
}

\markboth{Journal of \LaTeX\ Class Files, Vol. xx, No. xx, September 202x}
{Shell \MakeLowercase{\textit{et al.}}: Bare Demo of IEEEtran.cls for IEEE Journals}
\maketitle
\begin{abstract}
The present paper proposes a data-driven sensor selection method for a high-dimensional nondynamical system with strongly correlated measurement noise. The proposed method is based on proximal optimization and determines sensor locations by minimizing the trace of the inverse of the Fisher information matrix under a block-sparsity hard constraint. The proposed method can avoid the difficulty of sensor selection with strongly correlated measurement noise, in which the possible sensor locations must be known in advance for calculating the precision matrix for selecting sensor locations. The problem can be efficiently solved by the alternating direction method of multipliers, and the computational complexity of the proposed method is proportional to the number of potential sensor locations when it is used in combination with a low-rank expression of the measurement noise model. The advantage of the proposed method over existing sensor selection methods is demonstrated through experiments using artificial and real datasets.
\end{abstract}


\IEEEpeerreviewmaketitle

\section{Introduction}
\IEEEPARstart{M}{easurements} of physical quantities are essential in various fields. Typically, measurements are made at specific points using discretely installed sensors. Therefore, it is necessary to maximize the information obtained with as few sensors as possible. This kind of situation can be seen in various types of measurements, such as global positioning system \cite{kihara1984satellite,phatak2001recursive}, acoustic measurements \cite{douganccay2009optimal,macho2016optimal}, structural health monitoring \cite{worden2001optimal,yi2011optimal}, environment monitoring \cite{du2014optimal}, brain source localization \cite{yeo2022efficient}, etc. 


The problem that optimizes sensor locations is called the sensor selection/placement problem and is formulated as a combinatorial optimization problem, which is known as an NP-hard problem. Table.~\ref{tab:methods} shows typical methods used in a sensor selection problem.
Global optimization techniques, such as branch and bound \cite{welch1982branch,lawler1966branch}, can obtain an exact solution to the sensor selection problem. However, because their computational costs are high, global optimization techniques can only be used for small (choosing a small number of sensors from a small number of potential sensor locations) problems. Therefore, convex relaxation and greedy methods, which can obtain a suboptimal solution with less computational time, have been studied.

Joshi and Boyd proposed \cite{joshi2009sensor} a sensor optimization method based on convex optimization. Their method can obtain a global optimal solution to the relaxed problem. The computational complexity of their method is proportional to the cube of the number of potential sensor locations, and thus, the computational cost is much lower than that of global optimization techniques. The accelerated randomized convex relaxation method was recently proposed \cite{nonomura2021randomized}, but the computational cost is still high for high-dimensional data, such as surface measurement or volume measurement data. 
A convex optimization method based on a proximal splitting algorithm is another sensor selection method. The sparsity-promoting framework was introduced by Fardad et al. \cite{fardad2011sparsity} and Lin et al. \cite{lin2013design}. Their framework allows us to obtain block-sparse feedback and observer gains, as well as select actuators and sensors in dynamical systems, and Dhingra et al. \cite{dhingra2014admm} and Zare and Jovanovi\'{c} \cite{zare2018optimal} extended those methods. 
\reviewerA{Masazade et al. \cite{masazade2012sparsity} proposed the method for selecting sparse sensors that minimize the estimation using an extended Kalman filter. Their method obtains the sparse Kalman gain matrix using the sparsity-promoting penalty function. In addition, sensor selection methods for a field estimation on a nondynamical system based on convex or nonconvex proximal optimization are available \cite{liu2014sparsity,nagata2021data}.}

Greedy methods provide a suboptimal solution to sensor selection problems at a low computational cost.
Manohar et al. \cite{manohar2018data} proposed a greedy method based on QR decomposition. Their method is related to the discrete empirical interpolation method \cite{chaturantabut2010nonlinear} and the QR-based discrete empirical interpolation method \cite{drmac2016new} in the framework of the Galerkin projection \cite{rowley2004model}. 
Saito et al. \cite{saito2021determinantbased} proposed a greedy method based on the D-optimal design of experiments.
They also constructed a sensor selection framework for multicomponent measurements \cite{saito2020data,saito2021data}.
Nakai et al. \cite{nakai2021effect} formulated a greedy method based on the E- and A-optimal design of experiments and investigated the influence of the objective function on the performance of the selected sensor subset. Greedy methods are extended to a further generalized form \cite{clark2018greedy,manohar2021optimal,saito2020data,saito2021data,yamada2021fast,yamada2022greedy,jiang2019group,clark2020multi,clark2020sensor,nakai2021effect,li2021efficient,nakai2022nondominated,nagata2022randomized}. However, it is difficult to apply the greedy method to complex objective functions and constraints.
\begin{table}
\caption{Characteristics of practical algorithms for sensor selection problems.}
\setlength{\tabcolsep}{3pt}
\begin{tabular}{|p{35pt}|p{65pt}|p{60pt}|p{55pt}|}
\hline
Method& 
Global optimization&
Convex relaxation&
Greedy method\\
\hline
Solution& 
Exact&
Global optimum&
Local optimum\\
Cost& 
Very expensive&
Expensive&
Cheap\\
Obj. func.& 
-&
Flexible&
Inflexible\\
\hline
\end{tabular}
\label{tab:methods}
\end{table}

In most cases, including the methods mentioned above, the observed noise is assumed to be white noise, but the measured data are often corrupted by correlated noise. Therefore, the development of a sensor selection method for correlated measurements is a critical task. Several researchers have studied sensor selection in the presence of correlated measurement noise. Liu et al. \cite{liu2016sensor} and Yamada et al. \cite{yamada2021fast} modified the formulations of the sensor selection problem by weighting the value of each sensor based on prior noise information. They employed a Bayesian estimation framework and also used prior information in the estimation. Ucinski et al. \cite{ucinski2020d} and O'Connor et al. \cite{o2016distributed} employed a similar formulation, which includes a weighting term based on prior noise information. In their framework, an ordinary least-squares estimation was used. The difficulty of treating correlated noise in the sensor selection problem is that the influence of noise cannot be evaluated unless the sensor location is determined. \reviewerA{The previous methods considering the correlated measurement noise require the calculation of the precision matrix (inverse of the noise covariance matrix) which should be constructed by the noise information over only selected sensors. Particularly, the method based on convex relaxation requires higher calculation cost or a complicated formulation with limitations. There is no published method based on the continuous optimization problem for sensor selection that can be solved by a method with a computational complexity less than a cubic order of the number of potential sensor locations. Hence, it is difficult to employ the previously proposed methods for large-scale sensor selection problems including such correlated measurement noise.}


In the present study, we consider a sensor selection problem involving selecting $p$ sensors from $n$ potential sensors in a high-dimensional nondynamical system with correlated measurement noise. The sensor subset gives an observation vector of a linear function of latent variables superimposed with spatially correlated measurement noise. We extended the sensor selection method based on the proximal splitting algorithm \cite{nagata2021data} to a problem including correlated measurement noise. 
Our goal is to choose a suboptimal sensor subset based on the optimal design of experiments considering the influence of correlated measurement noise. The main contributions of the present study are as follows:
\begin{itemize}
    \item Sensor selection based on the proximal splitting algorithm for nondynamical systems with strongly correlated measurement noise is proposed.
    \item The proposed method can avoid the difficulty (described in Section~\ref{sec:diff}) of sensor selection with strongly correlated measurement noise in which the sensor locations must be known in advance for selecting the sensor locations. The problem can be efficiently solved by the alternating direction method of multipliers (ADMM) \cite{gabay1976dual,eckstein1992douglas}, and the computational complexity of the proposed method is $\mathcal{O}\left(n\right)$ using a low-rank expression of the measurement noise model \cite{yamada2021fast,yamada2022greedy}. 
    \item The proposed method is applied to artificial and practical datasets, the National Oceanic and Atmospheric Administration Optimum Interpolation Sea Surface Temperature (NOAA-OISST) dataset. The optimization results are compared with those of the greedy method.
    \item Although the computational time of the proposed method is longer than that of the greedy method, the proposed method is better than the previously proposed greedy methods in terms of objective value and reconstruction error.
\end{itemize}
We laid the groundwork for further extension to complex objective functions, including constraints. MATLAB code for our algorithm can be found at https://github.com/Aerodynamics-Lab/Proximal-Optimization-based-Sensor-Selection-Algorithm-with-Correlated-Measurement-Noise \cite{nagata2022github}.

\section{Sensor Selection Problem}
\reviewerB{A mathematical description of the problem and the proposed method is shown in this section. The notation table that summarizes the main mathematical symbols and their definitions is shown in Table~\ref{tab:notation}.}

\begin{table}
\caption{\reviewerB{Variables and operators}}
\centering
\setlength{\tabcolsep}{3pt}
\begin{tabular}{p{20pt}p{200pt}}
\hline
\multicolumn{1}{c}{\reviewerB{Symbol}} & \multicolumn{1}{c}{\reviewerB{Description}} \\
\hline
\reviewerB{$\mathbf{C}$} & \reviewerB{Measurement matrix}\\
\reviewerB{$\mathbf{H}$} & \reviewerB{Sensor location matrix} \\
\reviewerB{$\mathbf{I}$} & \reviewerB{Identity matrix} \\
\reviewerB{$\mathbf{K}$, $\mathbf{X}$} & \reviewerB{Gain matrix} \\
\reviewerB{$\mathbf{R}$, $\mathbf{Q}$} & \reviewerB{Noise covariance matrix} \\
\reviewerB{$\mathbf{R}_{\rm d}$} & \reviewerB{Diagonal matrix with the diagonal entries of $\mathbf{R}$} \\
\reviewerB{$\mathbf{U}$} & \reviewerB{Sensor candidate matrix} \\
\reviewerB{$\mathbf{x}$} & \reviewerB{Snapshot of full state} \\
\reviewerB{$\mathbf{y}$} & \reviewerB{Observation vector} \\
\reviewerB{$\mathbf{z}$} & \reviewerB{Latent variables} \\
\hline
\multicolumn{1}{c}{\reviewerB{Symbol}} & \multicolumn{1}{c}{\reviewerB{Description}} \\
\hline
\reviewerB{$\mathbf{A}^{\dagger}$} & \reviewerB{Moore--Penrose pseudo inverse of $\mathbf{A}$} \\
\reviewerB{$\hat{\mathbf{A}}$} & \reviewerB{Normalized $\mathbf{A}$ by noise weighting term $\mathbf{R}_{\rm d}^{-\frac{1}{2}}$, i.e., $\hat{\mathbf{A}}=\mathbf{R}_{\rm d}^{-\frac{1}{2}}\mathbf{A}$} \\
\reviewerB{$\mathrm{tr}{\left(\mathbf{A}\right)}$} & \reviewerB{Trace norm of $\mathbf{A}$, i.e., $\mathrm{tr}{\left(\mathbf{A}\right)}=\Sigma_i \mathrm{diag}_i\left(\mathbf{A}\right)$} \\
\reviewerB{$\mathbf{A}^\mathsf{T}$} & \reviewerB{Transposition} \\
\reviewerB{$\mathbf{A}_{b:c}$} & \reviewerB{Truncated matrix of $\mathbf{A}$ from $b$-th column/row to $c$-th column/row} \\
\reviewerB{$\|\mathbf{A}\|_{\rm F}$} & \reviewerB{Frobenius norm of $\mathbf{A}$, i.e., $\|\mathbf{A}\|_{\rm F}=\sqrt{\Sigma_i\Sigma_j \|a_{ij}\|^2}$} \\
\reviewerB{$\|\mathbf{A}\|_0$} & \reviewerB{Group $\ell_0$ pseudo-norm $\mathbf{A}$, i.e., $\|\mathbf{A}\|_{g,0}:=\|\left(\|\mathbf{a}_1\|_2,\cdots,\|\mathbf{a}_n\|_2\right)\|_0$} \\
\reviewerB{$\|\mathbf{a}\|_1$} & \reviewerB{Manhattan ($\ell_1$) norm of $\mathbf{a}$, i.e., $\|\mathbf{a}\|_1=\Sigma_i |\mathbf{a}_i|$} \\
\reviewerB{$\|\mathbf{a}\|_2$} & \reviewerB{Euclidean ($\ell_2$) norm of $\mathbf{a}$, i.e., $\|\mathbf{a}\|_2=\sqrt{\Sigma_i \mathbf{a}^2_i}$} \\
\reviewerB{$\mathbf{a}_i$} & \reviewerB{$i$-th row or column vector of $\mathbf{A}$}\\
\reviewerB{$\tilde{\mathbf{a}}$} & \reviewerB{Estimated $\mathbf{a}$} \\
\hline
\end{tabular}
\label{tab:notation}
\end{table}

\subsection{Problem Formulation}
A snapshot measurement of the full state $\mathbf{x}\in\mathbb{R}^n$ including a noise signal $\mathbf{v}\sim\mathcal{N}(\mathbf{0},\sigma^2\mathbf{I}) \in\mathbb{R}^p$ through sparse sensors can be expressed as follows:
\begin{align}
    \mathbf{y}=\mathbf{H}\mathbf{x+v},
    \label{eq:observation_org}
\end{align}

\noindent where $\mathbf{y}\in\mathbb{R}^p$ and $\mathbf{H}\in\mathbb{R}^{p\times n}$ are the observation vector and the sensor location matrix, respectively. Here, $n$ and $p$ are the numbers of potential sensors and activated sensors, respectively. The sensor location matrix $\mathbf{H}$ is a sparse matrix, and each row vector of $\mathbf{H}$ is a unit vector. Unity locations in the row vectors correspond to the activated sensor locations chosen from $n$ potential sensor locations. Equation~\eqref{eq:observation_org} can be rewritten as follows using the sensor candidate matrix $\mathbf{U}\in\mathbb{R}^{n\times r}$ and the latent variables $\mathbf{z}\in\mathbb{R}^r$:
\begin{align}
    \mathbf{y}&\approx\mathbf{H}\mathbf{U}\mathbf{z+v} \nonumber \\
              &=\mathbf{C}\mathbf{z+v}.
    \label{eq:observation}
\end{align}

\noindent This system represents the problem of choosing $p$ sensors for the observation of signals generated by $r$ latent variables. The matrix $\mathbf{C}\in\mathbb{R}^{p\times r}$ is the measurement matrix, which is the product of the sensor location matrix and the sensor candidate matrix. Because $n$ corresponds to the degrees of freedom in the spatial direction of the full state, the sensor candidate matrix $\mathbf{U}$ is usually a tall-and-skinny matrix (i.e., $n>>r$) in a practical problem. The latent variables at $p>r$ can be estimated using the least-squares method using a pseudo-inverse operation,
\begin{align}
    \mathbf{\tilde{z}}
    &    =\mathbf{C}^{\dagger}\mathbf{y} \nonumber \\ 
    &=\left(\mathbf{C}^{\mathsf{T}}\mathbf{C}\right)^{-1}\mathbf{C}^{\mathsf{T}}\mathbf{y},
    \label{eq:LS_estimation}
\end{align}

\noindent where $\mathbf{\tilde{z}}$ is the set of estimated latent variables. The formulation in $p<r$ is not introduced due to the constraint used in the present formulation. 
Descriptions of the estimation of latent variables in $p<r$ are presented in \cite{yamada2022greedy}.

\reviewerA{In the case of measured signal including correlated measurement noise 
$\mathrm{E}\left[\mathbf{v}_{\mathrm c}\mathbf{v}_{\mathrm c}^\mathsf{T}\right]=\mathbf{R}_p$}, ``whitening'' of the measurement $\mathbf{y}$ is required before conducting the least-squares estimation as follows \cite{yamada2022greedy}:
\begin{align}
    \mathbf{R}_p^{-\frac{1}{2}}\mathbf{y}=\mathbf{R}_p^{-\frac{1}{2}}\mathbf{Cz}+\mathbf{R}_p^{-\frac{1}{2}}\mathbf{v}_\mathrm{c} \nonumber \\
    \Rightarrow~~\mathbf{y}_{\rm n}=\mathbf{C}_{\rm n}\mathbf{z}+\reviewerA{\mathbf{v}_{\rm n}},
\end{align}

\noindent where subscript $\circ_{\rm n}$ indicates the values corrected by the noise weighting term of the activated sensors $\mathbf{R}_p^{-\frac{1}{2}}\in\mathbb{R}^{p\times p}$, and $\mathbf{v}_{\rm n}$ is white noise. Therefore, 
$\mathrm{E}\left[\mathbf{v}_{\rm n}\mathbf{v}_{\rm n}^{\mathsf T}\right]=\sigma$ is a constant, and thus, the least-squares estimation of $\tilde{\mathbf{z}}$ with the whitened measurement vector $\mathbf{y}_{\rm n}$ at $p>r$ is obtained as follows:
\begin{align}
\tilde{\mathbf{z}}=\mathbf{C}_{\rm n}^\dagger\mathbf{y}_{\rm n}&=\left(\mathbf{C}_{\rm n}^{\mathsf{T}}\mathbf{C}_{\rm n}\right)^{-1}\mathbf{C}_{\rm n}^{\mathsf{T}}\mathbf{y}_{\rm n} \nonumber \\
&=\left(\mathbf{C}^{\mathsf{T}}\mathbf{R}_p^{-1}\mathbf{C}\right)^{-1}\mathbf{C}^{\mathsf{T}}\mathbf{R}_p^{-1}\mathbf{y}. \label{eq:WLS_estimation}
\end{align}

\subsection{A-optimality Criterion}
A-optimality criterion is the trace of the inverse of the Fisher information matrix (FIM). The A-optimal design minimizes the mean squared error in the estimation of the latent variables $\mathbf{\tilde{z}}$. Here, we consider the correlated measurement noise  $\mathbf{v}_{\rm c}$ in the observed signal of $\mathbf{x}$.
\begin{align}
    \mathbf{\tilde{z}}&=\left(\mathbf{C}^{\mathsf{T}}\mathbf{R}_p^{-1}\mathbf{C}\right)^{-1}\mathbf{C}^{\mathsf{T}}\mathbf{R}_p^{-1}\mathbf{y} \nonumber \\
                &=\left(\mathbf{C}^{\mathsf{T}}\mathbf{R}_p^{-1}\mathbf{C}\right)^{-1}\mathbf{C}^{\mathsf{T}}\mathbf{R}_p^{-1}\mathbf{C}\mathbf{z}+\left(\mathbf{C}^{\mathsf{T}}\mathbf{R}_p^{-1}\mathbf{C}\right)^{-1}\mathbf{C}^{\mathsf{T}}\mathbf{R}_p^{-1}\reviewerA{\mathbf{v}_{\rm c}}
    \label{eq:ztilde}
\end{align}

\noindent Therefore, the estimation error can be computed as follows:
\begin{align}
    \mathbf{\tilde z-z}&=\left(\mathbf{C}^{\mathsf T}\mathbf{R}_p^{-1}\mathbf{C}\right)^{-1}\mathbf{C}^{\mathsf T}\mathbf{R}_p^{-1}\mathbf{v}_{\reviewerA{\mathrm{c}}}.
\end{align}

\noindent To consider the average variance of the estimation, the trace norm of the error covariance matrix is calculated as follows:
\begin{align}
    &\mathrm{tr}\left(\mathrm{E}\left[\left(\mathbf{\tilde z}-\mathbf{z}\right)\left(\mathbf{\tilde z}-\mathbf{z}\right)^{\mathsf T}\right]\right) \nonumber \\
    &=\mathrm{tr}\left(\mathrm{E}\left[\left(\mathbf{C}^{\mathsf T}\mathbf{R}_p^{-1}\mathbf{C}\right)^{-1}\mathbf{C}^{\mathsf T}\mathbf{R}_p^{-1}\mathbf{v}_{\rm c}\mathbf{v}_{\rm c}^{\mathsf T}\mathbf{R}_p^{-1}\mathbf{C}\left(\mathbf{C}^{\mathsf T}\mathbf{R}_p^{-1}\mathbf{C}\right)^{-1}\right]\right).
\label{eq:esterr}
\end{align}

\noindent Thus, the sensor selection problem based on the A-optimality criterion can be expressed as a minimization problem of a following objective function:
\begin{align}
f=\mathrm{tr}\,\left(\left(\mathbf{C}^{\mathsf{T}}\mathbf{R}_p^{-1}\mathbf{C}\right)^{-1}\right).~~~~~(p\geq r)
\label{eq:obj_tr}
\end{align}

\subsection{Noise Covariance}
There is a situation in which the data include not only Gaussian noise but also correlated measurement noise. In data-driven sensor selection, the reduced-order modeling of a data matrix, which is given as the $r$-rank approximation \cite{eckart1936approximation} obtained by the singular value decomposition (SVD) $\mathbf{X}_\textrm{data}\approx\mathbf{U}_{1:r_1}\mathbf{S}_{1:r_1}\mathbf{V}_{1:r_1}^{\mathsf T}$, is used. Here, $\mathbf{X}_\textrm{data}\in\mathbb{R}^{n\times m}$ is training data matrix containing snapshots $\mathbf{X}_\textrm{data}=\left[\mathbf{x_1}\,\cdots\,\mathbf{x_m}\right]$. \reviewerA{The measurement noise consists of the truncated modes of from $(r_1+1)$th to $m$th modes because those components are included in the measurement but not in the reduced-order model.} In the present study, the spatial covariance derived from the truncated SVD modes of the training data matrix is used as the noise model, as in previous work based on the greedy method of Yamada et al. \cite{yamada2021fast,yamada2022greedy}. The snapshot with full-state observation $\mathbf{x}$ and noise covariance matrix $\mathbf{R}\in\mathbb{R}^{n\times n}$ become
\begin{align}
    \mathbf{x}=\mathbf{U}\mathbf{z}+\mathbf{w},
\end{align}

\noindent \reviewerA{where $\mathbf{w}$ is the correlated noise on the full-state observation and its covarionce is given by $\mathbf{R}$ as follows:}
\begin{align}
    \mathrm{E}\left[\mathbf{ww^{\mathsf T}}\right]&\reviewerD{=:} \mathbf{R}.
\end{align}

The observations through the sensors and noise covariance are
\begin{align}
    \mathbf{y}=\mathbf{HU}\mathbf{z}+\mathbf{Hw},
\end{align}

\noindent and
\begin{align}
    \mathrm{E}\left[\mathbf{Hww}^{\mathsf T}\mathbf{H}^{\mathsf T}\right]&= \mathbf{H}\mathrm{E}\left[\mathbf{ww}^{\mathsf T}\right]\mathbf{H}^{\mathsf T} \nonumber \\
    &\reviewerD{=}\mathbf{H}\mathbf{R}\mathbf{H}^{\mathsf T}\reviewerD{=:}\mathbf{R}_p, 
    \label{eq:Rp}
\end{align}

\noindent respectively. The matrix $\mathbf{R}_p\in\mathbb{R}^{p\times p}$ is the noise covariance matrix for selected sensors. The noise covariance is assumed to be represented by the truncated SVD modes from $r_1+1$ to $m$ of the training data matrix as follows:

\begin{align}
    \mathbf{R}&=\mathrm{E}\left(\mathbf{ww}^{\mathsf T}\right) \\
    &=\mathrm{E}\left(\left(\mathbf{x}-\mathbf{U}_{1:r_1}\mathbf{z}\right)\left(\mathbf{x}-\mathbf{U}_{1:r_1}\mathbf{z}\right)^{\mathsf T}\right) \nonumber \\
    &\approx\left(\mathbf{USV}^{\mathsf T}-\mathbf{U}_{1:r_1}\mathbf{S}_{1:r_1}\mathbf{V}_{1:r_1}^{\mathsf T}\right)\left(\mathbf{USV}^{\mathsf T}-\mathbf{U}_{1:r_1}\mathbf{S}_{1:r_1}\mathbf{V}_{1:r_1}^{\mathsf T}\right)^{\mathsf T} \\
    &=\left(\mathbf{U}_{(r_1+1):m}\mathbf{S}_{(r_1+1):m}\mathbf{V}_{(r_1+1):m}^{\mathsf T}\right)\left(\mathbf{U}_{(r_1+1):m}\mathbf{S}_{(r_1+1):m}\mathbf{V}_{(r_1+1):m}^{\mathsf T}\right)^{\mathsf T} \nonumber\\
    &=\mathbf{U}_{(r_1+1):m}\mathbf{S}_{(r_1+1):m}^2\mathbf{U}_{(r_1+1):m}^{\mathsf T}. \label{eq:ncov}
\end{align}

\noindent When higher-order modes can be ignored because their contributions are small, \eqref{eq:ncov} can be truncated, and a further low-dimensional form of the noise covariance matrix can be obtained as
\begin{align}
    \mathbf{R}\approx\mathbf{U}_{(r_1+1):r_2}\mathbf{S}_{(r_1+1):r_2}^2\mathbf{U}_{(r_1+1):r_2}^{\mathsf T} \label{eq:ncovlow},
\end{align}

\noindent where $r_2\leq m$. In the present study, the diagonal matrix with the difference between the original noise covariance matrix and the low-rank noise covariance matrix $\Delta \mathbf{S}\in\mathbb{R}^{n\times n}$ is computed as follows:
\begin{align}
    \reviewerCC{\Delta\mathbf{S}={\rm diag}\left({\rm diag}\left(\mathbf{R}\right)-{\rm diag}\left(\mathbf{U}_{(r_1+1):r_2}\mathbf{S}_{(r_1+1):r_2}^2\mathbf{U}_{(r_1+1):r_2}^{\mathsf T}\right)\right)\label{eq:deltas}.}
\end{align}

\noindent By adding this diagonal matrix, the approximation accuracy of the noise covariance matrix was improved.
\begin{align}
    \mathbf{R}\approx\mathbf{U}_{(r_1+1):r_2}\mathbf{S}_{(r_1+1):r_2}^2\mathbf{U}_{(r_1+1):r_2}^{\mathsf T}+\Delta \mathbf{S} \label{eq:ncovlow2}
\end{align}

\subsection{Difficulty in Treatment of Correlated Noise in Sensor Selection}\label{sec:diff}
The difficulty of treating correlated noise in sensor selection problems is that the influence of noise cannot be evaluated unless the sensor location is determined. The previous methods considering the correlated measurement noise require the calculation of the precision matrix (inverse of the noise covariance matrix) which should be constructed by the noise information over only selected sensors. As shown in \eqref{eq:obj_tr} and \eqref{eq:Rp}, the matrix $\mathbf{R}_p$ constructed as $\mathbf{HRH}^{\mathsf T}$ is required for evaluation of the objective function when selecting the sensor locations $\mathbf{H}$. Thus, the sensor location matrix $\mathbf{H}$ is required for determining the sensor locations. Therefore, consideration of the correlated measurement noise in the conventional convex relaxation method \cite{joshi2009sensor} is difficult. Previous studies \cite{rigtorp2010sensor,jamali2014sparsity,shen2014sensor} considered weakly correlated noise and approximated the FIM in the following form:
\begin{align}
    \mathbf{C}^{\mathsf T}\mathbf{R}_p^{-1}\mathbf{C} \approx \mathbf{F}=\mathbf{U}^{\mathsf T}\left(\mathbf{ss}^{\mathsf T}\circ\mathbf{R}^{-1}\right)\mathbf{U},
    \label{eq:weakcorr}
\end{align}

\noindent
where the variable $\mathbf{s}\in\mathbb{R}^n$ is the solution vector for the convex relaxation method with constraints ${\mathbf 1}^{\mathsf T}\mathbf{s}=p$ and $s_i\in(0,1)$, which correspond to the weights for the sensor candidates. The problem is solved by semidefinite programming.
The inverse of the noise covariance matrix (the precision matrix for observations) in this FIM is clearly affected by the noise information for the sensors that are not selected when taking the inverse of $\mathbf{R}$. This is because all the noise information can be observed, even though the signal can only be observed by the activated sensors in the system corresponding to this FIM. Therefore, this formulation can only be used for weakly correlated noise, as mentioned in \cite{liu2016sensor}.



Liu et al. \cite{liu2016sensor} avoided this difficulty by splitting the noise covariance matrix into a diagonal matrix $a\mathbf{I}$ and a matrix $\mathbf{B}$ as follows:
\begin{align}
    \mathbf{R}=a\mathbf{I}+\mathbf{B},
\end{align}
\noindent
where $a$ is chosen such that the matrix $\mathbf{B}$ is positive definite. In this case, the noise covariance matrix for selected sensors $\mathbf{R}_p$ can be obtained as follows:
\begin{align}
    \mathbf{R}_p&=\mathbf{H}\left(a\mathbf{I}+\mathbf{B}\right)\mathbf{H}^{\mathsf T} \nonumber \\
                &=a\mathbf{I}+\mathbf{H}\mathbf{B}\mathbf{H}^{\mathsf T},
\end{align}
\noindent
and the approximated FIM $\mathbf{F}$ \eqref{eq:weakcorr} becomes
\begin{align}
    \mathbf{F}&=\mathbf{U}^{\mathsf T}\mathbf{H}^{\mathsf T}\mathbf{R}^{-1}\mathbf{H}\mathbf{U} \nonumber \\
    &=\mathbf{U}^{\mathsf T}\mathbf{B}^{-1}\mathbf{U}-\mathbf{U}^{\mathbf T}\mathbf{B}^{-1}\left(\mathbf{B}^{-1}+a^{-1}\mathrm{diag}\left(\mathbf{s}\right)\right)^{-1}\mathbf{B}^{-1}\mathbf{U}, \label{eq:liuFIM}
\end{align}
\noindent
where again the variable $\mathbf{s}\in\mathbb{R}^n$ is the solution vector for the convex relaxation method with constraints ${\mathbf 1}^{\mathsf T}\mathbf{s}=p$ and $s_i\in(0,1)$, which corresponds to the weights of the sensor candidates. This operation of separating nondiagonal components in $\mathbf{R}$ results in an objective function whose covariance in measurement noise is properly evaluated. The objective function based on \eqref{eq:liuFIM} can be solved by semidefinite programming. However, there is no published method to formulate the continuous optimization problem for sensor selection that can be solved by a method with a computational complexity less than $\mathcal{O}\left(n^3\right)$. We will introduce a formulation that can consider correlated noise without computing the precision matrix and that can solve the problem with a computational cost of $\mathcal{O}\left(n\right)$ in the following section.

\section{Proposed Method} \label{sec:prob}
\subsection{Mathematical Formulation}
The mathematics of the present method is based on \cite{nagata2021data}. The difference between the previous and present methods is whether to consider independent or correlated noise. Consider a matrix $\mathbf{K}_p\in\mathbb{R}^{r_1\times p}$ that recovers $\mathbf{z}$ from the observation $\mathbf{y}$ in \eqref{eq:LS_estimation}
\begin{align}
    \tilde{\mathbf{z}}&=\mathbf{K}_p\mathbf{y} \nonumber\\ 
                      &=\mathbf{KUz}+\mathbf{K\reviewerA{w}}\label{eq:ztildeK}.
\end{align}

\noindent
The gain matrix $\mathbf{K}\in\mathbb{R}^{r_1\times n}$ is a sparse matrix, which has $p$ nonzero column vectors. The locations of activated sensors are indicated by that of nonzero column vectors in the gain matrix $\mathbf{K}$. As in the previous study \cite{nagata2021data}, the gain matrix $\mathbf{K}$ is only used to construct the sensor matrix $\mathbf{H}$ in the polishing step (see Section~\ref{sec:polishing}) and is not used directly for the estimation of $\tilde{\mathbf{z}}$. 
In addition, $\tilde{\mathbf{z}}$ is assumed to be an unbiased approximation of $\mathbf{z}$. In this case,
\begin{align}
    \mathbf{KU}&=\mathbf{I}
\end{align}

\noindent 
should be satisfied. \reviewerC{This constraint is important for the present algorithm, and the optimization problem of the objective function with several constraints should be solved. Although the faster algorithm of the proximal gradient method, such as the fast iterative shrinkage thresholding algorithm \cite{beck2009fast} potentially seems to be a candidate, it cannot be applied to problems with a nondifferentiable objective function with constraints. Thus, ADMM that can handle the constraint with proximal operator is employed, as described in \ref{sec:ADMM}}. The average error in $\tilde{\mathbf{z}}$ can be explained in the same way as described in \eqref{eq:ztilde}.
\begin{align}
    &\mathrm{tr}\left(\mathrm{E}\left[\left(\mathbf{\tilde z}-\mathbf{z}\right)\left(\mathbf{\tilde z}-\mathbf{z}\right)^{\mathsf T}\right]\right) \nonumber \\
   &=\mathrm{tr}\left(\mathrm{E}\left[\mathbf{Kww}^{\mathsf T}\mathbf{K}^{\mathsf T}\right] \right)=\mathrm{tr}\left(\mathbf{KRK}^{\mathsf T} \right)
\label{eq:esterrD}
\end{align}

\noindent To obtain the sparse gain matrix $\mathbf{K}$, the group-sparsity paradigm \cite{yuan2006model} is introduced, and \eqref{eq:esterrD} is augmented with a sparsity-promoting term. 
When the group $\ell_1$-norm penalty is used as a sparsity-promoting term, the objective function of the sensor selection problem can be obtained as follows:
\begin{align}
    &\underset{\reviewerB{\mathbf{K}}}{\mathrm{minimize}}~\mathrm{tr}\left(\mathbf{KRK}^{\mathsf T}\right)+\lambda \sum_{i=1}^n\left|\left|\mathbf{k}_i\right|\right|_2~~\mathrm{subject~to}~\mathbf{KU=I},
    \label{eq:objorgGL1f}
\end{align}

\noindent
where ${\mathbf k}_i\in\mathbb{R}^n$ is the $i$th column vector of $\mathbf{K}$, and $\lambda>0$ is the regularization parameter that adjusts the sparsity of the solution.
The sparsity-promoting term penalizes a greater number of nonzero columns in $\mathbf{K}$, which corresponds to the number of selected sensors. The advantage of the present method is the flexibility of the objective function. In the case of the conventional method considering correlated noise, the covariance matrix constructed based on selected sensors, as discussed in Section \ref{sec:diff}, is required for evaluating the objective function for sensor selection. In contrast, the present method can minimize the objective function \eqref{eq:objorgGL1f} directly.

However, when sensor selection is performed using the objective function of \eqref{eq:objorgGL1f}, the number of sensors to be selected cannot be determined in advance due to the nature of the sparsity-promoting term. In the present study, therefore, a group $\ell_0$ pseudo-norm constraint is used to obtain $p$ sensors that determined by users in advance:
\begin{align}
    &\underset{\reviewerB{\mathbf{K}}}{\mathrm{minimize}}~\mathrm{tr}\left(\mathbf{KRK}^{\mathsf T}\right) \nonumber \\ 
    &\mathrm{subject~to}~\left|\left|\left(\left|\left|\mathbf{k}_1\right|\right|_2,\cdots,\left|\left|\mathbf{k}_n\right|\right|_2\right)\right|\right|_0\leq p~\mathrm{and}~\mathbf{KU=I}.
    \label{eq:objorgGL0fnorm}
\end{align}

\noindent
The operator $\left|\left|\,\cdot\,\right|\right|_0$ indicates the $\ell_0$ pseudo-norm, which returns the total number of nonzero entries, and the operator $\left|\left|\,\cdot\,\right|\right|_2$ indicates the $\ell_2$ norm of a vector. Although it does not lead to essential changes in the equations, a simple expression consistent with the standard problems adopted in ADMM can be obtained by taking the transpose of this formulation and the variable transformation:
\begin{align}
\left\{
\begin{matrix}
\mathbf{X} &=& \mathbf{K}^{\mathsf T}\\
\mathbf{A} &=& \mathbf{U}_{1:r_1}^{\mathsf T}\\
\Delta \mathbf{S_Q}&=&\Delta \mathbf{S} \\
\mathbf{U_Q}&=&\mathbf{U}_{(r_{1}+1):r_{2}} \\
\mathbf{S_Q}&=&\mathbf{S}_{(r_{1}+1):r_{2}} \\
\reviewerCC{\mathbf{Q}} &=&\reviewerCC{\mathbf{U_Q S_Q U_Q}^{\mathsf T}+\Delta \mathbf{S_Q}}. \\
\end{matrix}
\right.
\end{align}

\noindent
Then, the objective function can be written as follows:
\begin{align}
    &\underset{\reviewerB{\mathbf{X}}}{\mathrm{minimize}}~\mathrm{tr}\left(\mathbf{X}^{\mathsf T}\mathbf{QX}\right) \nonumber \\
    &\mathrm{subject~to}~\left|\left|\left(\left|\left|\mathbf{x}_1\right|\right|_2,\cdots,\left|\left|\mathbf{x}_n\right|\right|_2\right)\right|\right|_0\leq p~\mathrm{and}~\mathbf{AX=I},
    \label{eq:obj}
\end{align}

\noindent
where ${\mathbf x}_i\in\mathbb{R}^n$ is the $i$th row vector of $\mathbf{X}$. The matrix $\mathbf{Q}$ is the noise covariance matrix described by a low-rank expression, as given in \eqref{eq:ncovlow2}.



\subsection{Consideration of Noise Intensity in Thresholding Operations} \label{sec:normalization}
In the present study, the sensor candidate matrix $\mathbf{U}$ was normalized according to the noise weighting term $\mathbf{R}_{\rm d}^{-\frac{1}{2}}$ where $\mathbf{R}_{\rm d}\in\mathbb{R}^{n\times n}$ is a diagonal matrix with diagonal entries of the noise covariance matrix $\mathbf{R}$. The normalized sensor candidate matrix $\hat{\mathbf{U}}$ is generated as follows:
\begin{align}
    \hat{\mathbf{U}}&=\mathbf{R}_{\rm d}^{-\frac{1}{2}}\mathbf{U}.
    \label{eq:RdnormU}
\end{align}

\noindent
Therefore, the gain matrix $\mathbf{K}$ is
\begin{align}
    \hat{\mathbf{K}}&=\mathbf{K}\mathbf{R}_{\rm d}^{\frac{1}{2}},
\end{align}

\noindent
and the objective function of \eqref{eq:objorgGL0fnorm} becomes
\begin{align}
    &\underset{\reviewerB{\hat{\mathbf{K}}}}{\mathrm{minimize}}~\mathrm{tr}\left(\hat{\mathbf{K}}\mathbf{R}_{\rm d}^{-\frac{1}{2}}\mathbf{R}\mathbf{R}_{\rm d}^{-\frac{1}{2}}\hat{\mathbf{K}}^{\mathsf T}\right) \nonumber \\ 
    &\mathrm{subject~to}~\left|\left|\left(\left|\left|\hat{\mathbf{k}}_1\right|\right|_2,\cdots,\left|\left|\hat{\mathbf{k}}_n\right|\right|_2\right)\right|\right|_0\leq p~\mathrm{and}~\hat{\mathbf{K}}\hat{\mathbf{U}}=\mathbf{I}.
    \label{eq:objGL0fnorm}
\end{align}

\noindent
The objective function, which has the same form as \eqref{eq:obj}, can be obtained using the following variables:
\begin{align}
\left\{
\begin{matrix}
\reviewerCC{\mathbf{Q}}&=&\reviewerCC{\Delta \mathbf{S} + \mathbf{U}_{(r_{1}+1):r_{2}}\mathbf{S}_{(r_{1}+1):r_{2}}\mathbf{U}_{(r_{1}+1):r_{2}}^{\mathsf T}}\\
\reviewerCC{\mathbf{R_{\rm d}}}&=&\reviewerCC{{\rm diag}(\mathbf{Q})} \\
\reviewerCC{\Delta \mathbf{S_Q}}&=&\reviewerCC{\mathbf{R}_{\rm d}^{-\frac{1}{2}}\Delta\mathbf{S}\mathbf{R}_{\rm d}^{-\frac{1}{2}}} \\
\reviewerCC{\mathbf{X}} &=& \reviewerCC{\hat{\mathbf{K}}^{\mathsf T}(=\mathbf{K}\mathbf{R}_{\rm d}^{\frac{1}{2}})}\\
\reviewerCC{\mathbf{A}} &=& \reviewerCC{\hat{\mathbf{U}}_{1:r_1}^{\mathsf T}(=\mathbf{R}_{\rm d}^{-\frac{1}{2}}\mathbf{U})}\\
\reviewerCC{\mathbf{U_Q}}&=&\reviewerCC{\mathbf{R}_{\rm d}^{-\frac{1}{2}}\mathbf{U}_{(r_{1}+1):r_{2}}} \\
\reviewerCC{\mathbf{S_Q}}&=&\reviewerCC{\mathbf{S}_{(r_{1}+1):r_{2}}}. \\
\end{matrix}
\right.
\end{align}

\noindent
Then, the problem can be solved by the same procedure as described in Section~\ref{sec:ADMML0}. It should be noted that the solution of the problem, the gain matrix $\mathbf{K}$, is obtained by denormalizing the solution matrix $\mathbf{X}$ as follows:
\begin{align}
\mathbf{K}=\mathbf{X}^{\mathsf T}\mathbf{R}_{\rm d}^{-\frac{1}{2}}.
\end{align}

\noindent
Normalization is required to consider the intensity of the noise in the proximal operator used in ADMM. In the case of the objective function \eqref{eq:objorgGL0fnorm}, the noise covariance matrix is considered in the first term of the objective function, but it is not considered in the sparsity-promoting term. This is because the effect of the correlated noise cannot be accurately evaluated unless the sensor locations are determined, as discussed in Section~\ref{sec:diff}. Therefore, it is difficult to consider the correlated measurement noise in the sparsity-prompting term. 
In the present method, the sensor candidate matrix $\mathbf{U}$ was normalized based on the diagonal entries of the noise covariance matrix, but off-diagonal entries cannot be considered. By normalization of the sensor candidate matrix based on the noise weighting term, the diagonal entries of the noise covariance matrix are considered in the computation of the sparsity-promoting term, even though the influence of the off-diagonal entries is still not considered. It should be noted that the noise considered in the present study has a structure of a certain size because the noise model consists of truncated modes. Hence, not only uncorrelated noise but also a part of the component of correlated noise will be considered by normalization using the noise weighting term, even if only the diagonal component of the noise covariance matrix is considered.

\subsection{Alternating Direction Method of Multipliers}\label{sec:ADMM}
In the present study, the optimization problem is solved by ADMM. When the sparsity-promoting term is the simple group $\ell_1$-norm penalty, \eqref{eq:objorgGL1f} has the following form:
\begin{align}
    \underset{\reviewerB{\mathbf{X},\,\mathbf{Z}}}{\mathrm{minimize}}~~g(\mathbf{X})+h(\mathbf{Z})~~~~\mathrm{subject}~\mathrm{to}~~\mathbf{Z=GX}.
    \label{eq:ADMMform}
\end{align}

\noindent 
For arbitrarily chosen $\mathbf{Z}^{(0)}$, $\mathbf{Y}^{(0)}$, and $\gamma>0$, ADMM iterates the following steps:
\begin{align}
    \left\{
    \begin{array}{ll}
    \mathbf{X}^{(k+1)}&=\underset{\mathbf{X}}{\mathrm{argmin}}~g\left(\mathbf{X}\right)+\frac{1}{2\gamma}\left|\left|\mathbf{Z}^{(k)}-\mathbf{GX}-\mathbf{Y}^{(k)}\right|\right|_2^2  \\
    \mathbf{Z}^{(k+1)}&=\underset{\mathbf{Z}}{\mathrm{argmin}}~h\left(\mathbf{Z}\right)+\frac{1}{2\gamma}\left|\left|\mathbf{Z}-\mathbf{GX}^{(k+1)}-\mathbf{Y}^{(k)}\right|\right|_2^2 \\
    \mathbf{Y}^{(k+1)}&=\mathbf{Y}+\mathbf{GX}^{(k+1)}-\mathbf{Z}^{(k+1)}.
    \end{array}
    \right. \label{eq:ADMM}
\end{align}
\noindent
In the present formulation with the simple group $\ell_1$-norm penalty, $g(\mathbf{X})$, $h(\mathbf{Z})$, $\mathbf{Z}$, and $\mathbf{G}$ correspond to
\begin{align}
    g\left(\mathbf{X}\right):=\mathrm{tr}\left(\mathbf{X}^{\mathsf T}\mathbf{QX}\right),&~h\left(\mathbf{Z}\right):=\lambda \sum_i \left|\left|\mathbf{x}_i\right|\right|_2, \\
    \mathbf{Z}=\left[
    \begin{array}{cc}
         \mathbf{Z}_1\\ \mathbf{Z}_2
    \end{array}\right]
    ,~{\rm and} &~\mathbf{G}=\left[
    \begin{array}{cc}
        \mathbf{I}\\ \mathbf{A}
    \end{array}\right],  \nonumber
\end{align}

\noindent
respectively. 

\subsection{Sensor Selection With Group $\ell_0$ Pseudo-Norm Constraint} \label{sec:ADMML0}
\subsubsection{Reformulation of Objective Function \eqref{eq:obj}}
The first constraint in \eqref{eq:obj}, which is the sparsity-promoting term, was computed by an $\ell_0$-constrained block hard thresholding ($L_0$BHT) operator. This operator can determine the group $\ell_0$ pseudo-norm of the solution matrix beforehand that corresponds to the number of sensors to be activated. 
To establish the algorithm for solving the problem \eqref{eq:obj}, we first reformulate the objective function \eqref{eq:obj} into an ADMM-applicable form defined in \eqref{eq:ADMMform}. We also define the group $\ell_0$ pseudo-norm used in $L_0$BHT as follows:
\begin{definition}[Group $\ell_0$ pseudo-norm]
The activated sensor locations correspond to the locations of nonzero row vectors of $\mathbf{X}$, and thus, the group $\ell_0$ pseudo-norm of $\mathbf{X}$ used in \eqref{eq:obj} as the constraint is defined as
\begin{align}
\|\mathbf{X}\|_{g,\,0}:=\|\left(\|\mathbf{x}_1\|_2,\cdots,\|\mathbf{x}_n\|_2\right)\|_0.
\label{eq:GL0norm}
\end{align}

\noindent
Then, the objective function \eqref{eq:obj} becomes 
\begin{align}
    &\underset{\reviewerB{\mathbf{X}}}{\mathrm{minimize}}~\mathrm{tr}\left(\mathbf{X}^{\mathsf T}\mathbf{QX}\right) 
    &\mathrm{subject~to}~\|\mathbf{X}\|_{g,\,0}\leq p~\mathrm{and}~\mathbf{AX=I}.
    \label{eq:obj2}
\end{align}

\noindent
In addition, we define the indicator function of the inequality constraint on the group $\ell_0$ pseudo-norm:
\begin{align}
l_{\{\|\cdot\|_{g,\,0}\,\leq\,p\}}\left(\mathbf{X}\right):=
    \left\{
    \begin{array}{cc}
        0,&~~~\|\mathbf{X}\|_{g,\,0}\leq p, \\
        \infty,&~~~\mathrm{otherwise}. \\
     \end{array}
     \right.\label{eq:indL0}
\end{align}

\noindent
Then, the objective function \eqref{eq:obj2} can be reformulated as
\begin{align}
    &\underset{\reviewerB{\mathbf{X}}}{\mathrm{minimize}}~\mathrm{tr}\left(\mathbf{X}^{\mathsf T}\mathbf{QX}\right)+l_{\left\{\|\cdot\|_{g,\,0}\,\leq\,p\right\}}\left(\mathbf{X}\right)
    &\mathrm{subject~to}~\mathbf{AX=I}.
    \label{eq:obj3}
\end{align}

\subsubsection{Algorithm}
The form of \eqref{eq:obj3} is the same as that of \eqref{eq:ADMMform}
\begin{align}
    g\left(\mathbf{X}\right):=\mathrm{tr}\left(\mathbf{X}^{\mathsf T}\mathbf{QX}\right),&~h\left(\mathbf{Z}\right):=l_{\|\cdot\|_{g,\,0}\,\leq\,p}\left(\mathbf{Z}\right), \\
    \mathbf{Z}=\left[
    \begin{array}{cc}
         \mathbf{Z}_1\\ \mathbf{Z}_2
    \end{array}\right]
    ~{\rm and} &~\mathbf{G}=\left[
    \begin{array}{cc}
        \mathbf{I}\\ \mathbf{A}
    \end{array}\right]. \nonumber
\end{align}

\noindent
This can be solved by ADMM as described in \eqref{eq:ADMM}. For arbitrarily chosen $\mathbf{Z}^{(0)}$, $\mathbf{Y}^{(0)}$, and $\gamma>0$, the proposed algorithm iterates following subproblems:
\begin{align}
    \left\{
    \begin{array}{ll}
    \mathbf{X}^{(k+1)}&=\underset{\mathbf{X}}{\mathrm{argmin}}~\mathrm{tr}\left(\mathbf{X}^{\mathsf T}\mathbf{QX}\right)+\frac{1}{2\gamma}\left|\left|\mathbf{Z}^{(k)}-\mathbf{GX}-\mathbf{Y}^{(k)}\right|\right|_2^2 \\
    \mathbf{Z}^{(k+1)}&=\underset{\mathbf{Z}}{\mathrm{argmin}}~l_{\left\{\|\cdot\|_{g,\,0}\,\leq\, p\right\}}\left(\mathbf{Z}\right)+\frac{1}{2\gamma}\left|\left|\mathbf{Z}-\mathbf{GX}^{(k+1)}-\mathbf{Y}^{(k)}\right|\right|_2^2 \\
    \mathbf{Y}^{(k+1)}&=\mathbf{Y}+\mathbf{GX}^{(k+1)}-\mathbf{Z}^{(k+1)}.
    \end{array}
    \right.
\end{align}
\end{definition}

\noindent
The first subproblem is differentiable and convex, and thus, the solution can be characterized as follows:
\begin{align}
    \frac{\mathrm d}{{\mathrm d}{\mathbf X}}\left\{\mathrm{tr}\left(\mathbf{X}^{\mathsf T}\mathbf{QX}\right)+\frac{1}{2\gamma}\left|\left|\mathbf{Z}^{(k)}-\mathbf{GX}-\mathbf{Y}^{(k)}\right|\right|_2^2\right\}&=0 \nonumber \\
    2{\mathbf{QX}}+\frac{1}{\gamma}{\mathbf G}^{\mathsf T}{\mathbf G}{\mathbf X}+\frac{1}{\gamma}{\mathbf G^{\mathsf T}\left({\mathbf Y}^{(k)}-{\mathbf Z}^{(k)}\right)}&=0, 
\end{align}

\noindent
and the solution of the first subproblem can be obtained.
\begin{align}
    {\mathbf X}&=\left(2{\mathbf Q}+\frac{1}{\gamma}{\mathbf G}^{\mathsf T}{\mathbf G}\right)^{-1}\frac{1}{\gamma}{\mathbf G}^{\mathsf T}\left({\mathbf Z}^{(k)}-{\mathbf Y}^{(k)}\right) \nonumber \\ 
    &=\left(2{\mathbf Q}+\frac{1}{\gamma}{\mathbf I}+\frac{1}{\gamma}{\mathbf A}^{\mathsf T}{\mathbf A}\right)^{-1}\frac{1}{\gamma}\left\{\left({\mathbf Z_1}-{\mathbf Y_1}\right)+{\mathbf A}^{\mathsf T}\left({\mathbf Z_1}-{\mathbf Y_1}\right)\right\}
    \label{eq:solutionmatrix}
\end{align}

\noindent
\reviewerA{Here, the covariance matrix $\mathbf{Q}$ can be approximated using a low-rank form as described in \eqref{eq:ncovlow2}. In addition, the computational cost of the least-squares solution of the first subproblem can be reduced by adopting the inverse matrix lemma given below, when the number of latent variables is small}:
\begin{align}
    \left(2{\mathbf Q}+\frac{1}{\gamma}{\mathbf I}+\frac{1}{\gamma}{\mathbf A}^{\mathsf T}{\mathbf A}\right)^{-1}=\left(2{\mathbf U}_{\mathbf Q}{\mathbf S}^2_{\mathbf Q}{\mathbf U}^{\mathsf T}_{\mathbf Q}+\underbrace{2\Delta\mathbf{S}+\frac{1}{\gamma}{\mathbf I}+\frac{1}{\gamma}{\mathbf A}^{\mathsf T}{\mathbf A}}_{\mathbf J}\right)^{-1} \nonumber \\
    ={\mathbf J}^{-1}-{\mathbf J}^{-1}{\mathbf U}\left\{\left(2{\mathbf S^2}\right)^{-1}+{\mathbf U}^{\mathsf T}{\mathbf J}^{-1}{\mathbf U}\right\}^{-1}{\mathbf U}^{\mathsf T}{\mathbf J}^{-1},
\end{align}
\noindent
where
\begin{align}
    {\mathbf J}^{-1}&=\left(2\Delta{\mathbf S}+\frac{1}{\gamma}{\mathbf I}+\frac{1}{\gamma}{\mathbf A}^{\mathsf T}{\mathbf A}\right)^{-1} \nonumber \\
    &{\scriptstyle =\left(2\Delta{\mathbf S}+\frac{1}{\gamma}{\mathbf I}\right)^{-1}-\left(2\Delta{\mathbf S}+\frac{1}{\gamma}{\mathbf I}\right)^{-1}\frac{1}{\gamma}{\mathbf A}^{\mathsf T}\left({\mathbf I}+{\mathbf A}\left(2\Delta{\mathbf S}+\frac{1}{\gamma}{\mathbf I}\right)^{-1}\frac{1}{\gamma}{\mathbf A}^{\mathsf T} \right)^{-1}{\mathbf A}\left(2\Delta{\mathbf S}+\frac{1}{\gamma}{\mathbf I}\right)^{-1}. }
\end{align}

The present method reduces the computational cost using a low-rank expression of the correlated measurement noise \cite{yamada2022greedy}. Therefore, when a more general noise model in which the correlated measurement noise is inversely proportional to the distance between sensors is used, the computational cost becomes large.

By noticing the definition of the indicator function \eqref{eq:indL0}, the second subproblem, the update of $\mathbf{Z}$ can be rewritten as follows:
\begin{align}
\mathbf{Z}^{(k+1)}&=\underset{\mathbf{Z}}{\mathrm{argmin}}~\frac{1}{2\gamma}\left|\left|\mathbf{Z}-\mathbf{GX}^{(k+1)}-\mathbf{Y}^{(k)}\right|\right|_2^2  \nonumber\\ & \mathrm{subject~to}~{\|\mathbf{Z}\|_{g,\,0}\,\leq\,p},
\end{align}

\noindent
and this update can be rewritten by the proximal operator for the group $\ell_0$ pseudo-norm constraint,
\begin{align}
    \mathbf{Z}^{(k+1)}&=\mathrm{prox}_{l_{\left\{\|\cdot\|_{g,\,0}\,\leq\, p\right\}}}\left(\mathbf{GX}^{(k+1)}+\mathbf{Y}^{(k)}\right).
\end{align}

\noindent
The proximal operator of the indicator function $l_{\left\{\|\cdot\|_{g,\,0}\,\leq\,p\right\}}$ corresponds to metric projection onto the set satisfying the constraint $\|\mathbf{Z}\|_{g,\,0}\leq p$ \cite{ono2017l0},
\begin{align}
\mathrm{prox}_{l_{\left\{\|\cdot\|_{g,\,0}\,\leq\, p\right\}}}\left(\mathbf{V}\right)&=\underset{\mathbf{W}}{\mathrm{argmin}}\sum_{i=1}^n\left|\left|\mathbf{v}_i-\mathbf{w}_i\right|\right|^2_2\\&{\mathrm{subject~to}~\|\mathbf{W}\|_{g,\,0}\,\leq\,p} \nonumber \\
&=:P_{\left\{\|\cdot\|_{g,\,0}\,\leq\,p\right\}}\left(\mathbf{V}\right),
\end{align}

\noindent
where vectors $\mathbf{v}_i$ and $\mathbf{w}_i$ are row vectors of the matrices $\mathbf{V}$ and $\mathbf{W}$, respectively. The operation for this projection is almost the same as the block hard thresholding, but the threshold is $\left(\|{\mathbf v}_j\|_2\right)_p$, which is the $p$th largest value of $||{\mathbf v}_i||_2$: 
\begin{align}\label{eq:l0bht}
    P_{\left\{\|\cdot\|_{g,\,0}\,\leq\,p\right\}}\left({\mathbf V}\right):=
    \left\{
    \begin{array}{cc}
        {\mathbf v}_i,&~~~\|{\mathbf v}_i\|_2~\geq\left(\|{\mathbf v}_j\|_2\right)_p, \\
        0,&~~~\|{\mathbf v}_i\|_2~<\left(\|{\mathbf v}_j\|_2\right)_p. \\
     \end{array}
     \right.
\end{align}

\noindent

The procedures of the proposed algorithm are summarized in Algorithm~\ref{alg:propmeth}. 
\reviewerB{Because the problem is nonconvex when the hard thresholding operator is used, convergence is not guaranteed. ADMM may converge to a just local optimum, but it is expected that it will possibly have better convergence properties (e.g., faster convergence or convergence to a point with better objective value) than other local optimization methods. There is ample experimental evidence in the literature that supports the empirical convergence of ADMM, especially when the nonconvex problem at hand exhibits a ``favorable'' structure \cite{boyd2011distributed}. The present problem consists of the convex function and the nonconvex function. In general, this is hard to compute, but the nonconvex function of the present problem is the cardinality constraint, which keeps the $p$ largest magnitude elements and zeros out the rest. This is one of the special cases that can be exactly carried out.}
\reviewerB{In addition, a decreasing $\gamma$ strategy was employed and ADMM was stabilized for nonconvex optimization. A scalar $\eta$ is introduced to gradually decrease the value of $\gamma$. Similar strategies were adopted in existing methods using $\ell_0$-type norms \cite{cheng2014feature,storath2014jump,nguyen2015fast,matsuoka2017joint,ono2017l0}. This technique is supported by recent convergence analyses of ADMM for nonconvex cases (e.g., \cite{li2015global,hong2016convergence}). In these studies, the sequence generated by ADMM under appropriate conditions with sufficiently small $\gamma$ converges to a stationary point. Although these theoretical analyses employ overly strict assumptions and cover the problem formulations in many applications, including the present study, they provide insight into parameter settings.}

\reviewerC{The convergence and error bound of the nonconvex problem in the present study might be provided based on the Kurdyka-Lojasiewicz (KL) property, etc. The convergence of ADMM in nonconvex problems with certain properties has been analyzed by several researchers \cite{yang2017alternating,wang2019global}. However, we were unable to provide proof that our nonconvex problem satisfies the conditions that are required for convergence at present. The objective of the present paper is to provide the result that better sensor selection can be performed practically by the present ADMM-based method than the previous methods. Therefore, the theoretical analysis of convergence and the error bound are left for future researches.}

\begin{algorithm}
\caption{Proposed Algorithm for Sensor Selection With Group $\ell_0$ Pseudo-Norm Constraint (MATLAB code available online \cite{nagata2022github})} \label{alg:propmeth}
\begin{algorithmic}
\renewcommand{\algorithmicrequire}{\textbf{Input:}}
\renewcommand{\algorithmicensure}{\textbf{Output:}}
\REQUIRE $\mathbf{X}^{(0)}=\mathbf{A}^{-1}$, $\mathbf{Y}^{(0)}=\mathbf{Z}^{(0)}=0$, $p>0$, $\gamma>0$, and $0<\eta<1$
\ENSURE  $\mathbf{X}^{(k)}$
	\WHILE{ $\|\mathbf{X}^{(k)}-\mathbf{X}^{(k-1)}\|_{\mathrm F}>\epsilon$ }
    \STATE $\mathbf{X}^{(k+1)}=$
    \STATE {\scriptsize $\left({\mathbf J}^{-1}-{\mathbf J}^{-1}{\mathbf U}\left\{\left(2{\mathbf S^2}\right)^{-1}+{\mathbf U}^{\mathsf T}{\mathbf J}^{-1}{\mathbf U}\right\}^{-1}{\mathbf U}^{\mathsf T}{\mathbf J}^{-1}\right)\frac{1}{\gamma}\left\{\left({\mathbf Z_1}-{\mathbf Y_1}\right)+{\mathbf A}^{\mathsf T}\left({\mathbf Z_1}-{\mathbf Y_1}\right)\right\}$};
    \STATE $\mathbf{Z}^{(k+1)}=\mathrm{prox}_{l_{\left\{\|\cdot\|_{g,\,0}\,\leq\,p\right\}}}\left(\mathbf{GX}^{(k+1)}+\mathbf{Y}^{(k)}\right)$;
    \STATE $\mathbf{Y}^{(k+1)}=\mathbf{Y}+\mathbf{GX}^{(k+1)}-\mathbf{Z}^{(k+1)}$;
	\STATE $\gamma \leftarrow \eta\gamma$;
	\STATE $k \leftarrow k+1$;
	\ENDWHILE
\end{algorithmic}
\end{algorithm}

\subsection{Polishing Step} \label{sec:polishing}
Although latent variables $\mathbf{z}$ can be estimated directly by the gain matrix $\mathbf{K}$, the accuracy of the estimation is degraded to some extent by the sparsity-promoting terms. Therefore, the sensor location matrix $\mathbf{H}$ is constructed based on the gain matrix $\mathbf{K}$, and the gain is recalculated using \eqref{eq:WLS_estimation} rather than the gain matrix $\mathbf{K}$ directly. The corresponding entry in $\mathbf{H}$ is set to unity (which activates the corresponding sensor) when the $\ell_2$ norm of the corresponding column vector of $\mathbf{K}$ is nonzero (e.g., greater than $10^{-4}$). The latent variables $\tilde{\mathbf{z}}$ is estimated using \eqref{eq:WLS_estimation} after constructing $\mathbf{H}$.

\section{Results and Discussion}\label{sec:rand}
The performance of the proposed method was evaluated by applying it to an artificial dataset used in a previous study \cite{yamada2021fast}. The obtained results were compared to those obtained by the greedy methods considering white noise or correlated noise and the ADMM-based method considering white noise. Random data matrices, ${\mathbf X}_{\mathrm{data}}=\mathbf{USV}^{\mathsf T}$ (${\mathbf X}_{\mathrm{data}}\in\mathbb{R}^{n\times m}$), were generated, where $\mathbf{U}\in\mathbb{R}^{n\times m}$ and $\mathbf{V}\in\mathbb{R}^{m\times m}$ consist of $m$-orthogonal column vectors that were generated by QR decomposition of normally distributed [$\mathcal{N}(0,1)$] random matrix and diagonal entries of the diagonal matrix $\mathbf{S}\in\mathbb{R}^{m\times m}$ were $\mathrm{diag}\left(\mathbf{S}\right)=\left[1,1/\sqrt{2},1/\sqrt{3},...,1/\sqrt{m}\right]$, respectively. 

The leading-$r_1(=10)$ modes were retained for the construction of a reduced-order model of ${\mathbf X}_{\mathrm{data}}$, and a certain number of subsequent modes up to the $r_2$th mode (from the 11th to the 40th modes in the present study) were retained for the construction of the noise covariance matrix. The objective function is the A-optimality criterion for all methods. The greedy method considering white noise [Greedy(WN)]\reviewerCC{\cite{nakai2021effect}}, the greedy method considering correlated noise [Greedy(CN)], the ADMM-based method considering white noise [ADMM(WN)]\reviewerCC{\cite{nagata2021data}}, and the ADMM-based method considering correlated noise [ADMM(CN)] were tested.  \reviewerCC{Here, Greedy(CN) selects sensor locations by minimizing the objective function \eqref{eq:obj_tr} in each single-sensor subproblem. This method is the same as the reference \cite{yamada2022greedy}, except for the point that the objective function is based on the D-optimality criterion. In the case of Greedy(WN) \cite{nakai2021effect}, it is the same as Greedy(CN), except for the noise covariance matrix $\mathbf{R}_{\rm d}$ in the objective function \eqref{eq:obj_tr} is the identity matrix. In the case of ADMM(WN) \cite{nagata2021data}, it is the same as ADMM(CN), except for the noise covariance matrix $\mathbf{Q}$ in the objective function of ADMM(CN) is the identity matrix.}

\begin{figure}[!tbp]
\centerline{\includegraphics[width=9cm]{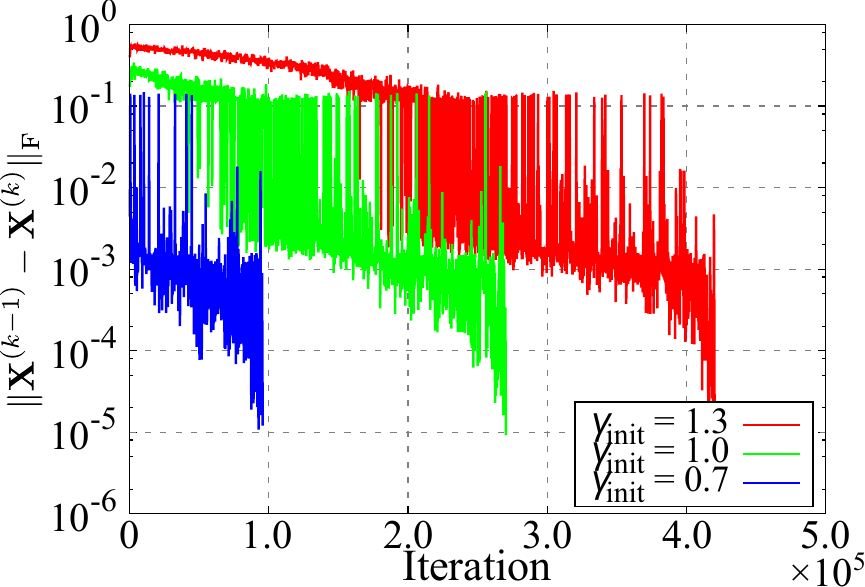}}
\caption{Iteration behavior for a certain case of ADMM(CN) for different $\gamma_\mathrm{init}$ for $p=30$.}
\label{fig:convergence} 
\end{figure}

In addition, ADMM(CNw/oN), which is the ADMM-based method without normalization by the noise weighting term described in Section~\ref{sec:normalization}, was also tested to investigate the effectiveness of normalization. The difference between ADMM(CN) and ADMM(CNw/oN) is described by the difference in their objective functions \eqref{eq:objGL0fnorm} and \eqref{eq:objorgGL0fnorm}. The computations using the ADMM-based methods were conducted with the $\ell_0$-constrained block hard thresholding ($L_0$BHT). \reviewerB{Because the problem is nonconvex when the hard thresholding operator is used, convergence is not guaranteed. Therefore, the decreasing $\gamma$ strategy was used. The step size $\gamma$ is initially set to be $\gamma=\gamma_\mathrm{init}$, and it was gradually reduced by multiplying by $\eta=0.99$ \cite{ono2017l0} every 5000 iterations. The iteration behavior of the present method [ADMM(CN)] for a certain case is shown in Figure~\ref{fig:convergence}. The size of dataset was ${\mathbf X}_{\mathrm{data}}\in\mathbb{R}^{10,000\times 100}$. The objective values obtained for $\gamma_\mathrm{init}=0.7$, 1.0, and 1.3 were $\mathrm{tr}\left(\mathbf{X}^\mathsf{T}\mathbf{RX}\right)=0.1526$, 0.1526, and 0.1517, respectively.}

In the following experiments, the value of $\gamma$ for ADMM-based methods was initially set at $\gamma_{\rm init}=1.0$. Each calculation in the numerical experiments was carried out 100 times with different datasets, and the computational time, the averaged objective value, and the reconstruction error were calculated.

\subsection{Computational Complexity and Computational Time}
The computational complexities of the previously proposed methods and the present method are summarized in Table.~\ref{tab:complexity}. \reviewerA{Here, SDP(CN) indicates the method based on semidefinite programming (SDP) proposed by Liu et al. \cite{liu2016sensor}. This method selects sensors while considering the influence of correlated measurement noise. Although prior knowledge (the matrix $\mathbf{\Sigma}$ in reference \cite{liu2016sensor}) can be considered when selecting sensor locations, it was not included in the present numerical experiment for a fair comparison.}

The number of operations in matrix multiplication and addition was evaluated based on the size of the matrices, whereas the matrix operations were accounted based on the Matlab code implementation. The largest complexity among each term was taken as the computational complexity of the algorithms. Here, the problem considered in the present study was $r_1\leq  p\approx (r_1-r_2)\ll n$.

\begin{table}
\caption{Comparison of Computational Complexity}
\setlength{\tabcolsep}{3pt}
\begin{tabular}{|p{125pt}|p{100pt}|}
\hline
Method& 
Computational complexity\\
\hline
Convex relaxation method(WN) \cite{joshi2009sensor}& 
${\mathcal O}\left(n^3\right)$ per iteration \\
\reviewerA{SDP(CN) \cite{liu2016sensor}}& 
\reviewerA{${\mathcal O}\left(n^{4.5}\right)$ per iteration} \\
Greedy(WN) \cite{nakai2021effect}& 
${\mathcal O}\left(pnr_1^2\right)$ \\
Greedy(CN) & ${\mathcal O}\left(pn(r_1-r_2)^2\right)$ \\
ADMM(WN) \cite{nagata2021data}& ${\mathcal O}\left(nr_1^2\right)$ per iteration \\ 
ADMM(CN) & ${\mathcal O}\left(n(r_1-r_2)^2\right)$ per iteration \\
\hline
\end{tabular}
\label{tab:complexity}
\end{table}

The computational complexities of Greedy(WN) and ADMM(WN) are $\mathcal{O}\left(pnr_1^2\right)$ and $\mathcal{O}\left(nr_1^2\right)$ (per iteration), respectively, where $p$ is the number of sensors to be activated, $n$ is the number of potential sensor locations, and $r_1$ is the number of latent variables. The computational complexity increases for methods that consider correlated noise. The computational complexities of Greedy(CN) and ADMM(CN) are $\mathcal{O}\left(pn\left(r_1-r_2\right)^2\right)$ and $\mathcal{O}\left(n\left(r_1-r_2\right)^2\right)$ (per iteration), respectively. \reviewerC{Regarding the computational complexity of each subproblem of the proposed method (Algorithm~\ref{alg:propmeth}), the complexity of the first subproblem is dominant, and it is $\mathcal{O}\left(n\left(r_1-r_2\right)^2\right)$. The complexities of the second and third subproblems are $\mathcal{O}\left(pn\right)$ and $\mathcal{O}\left(nr_1^2\right)$, respectively.} Although the computational cost of ADMM-based methods is independent of $p$, these methods require iterative computation, and thus, the characteristics of the dataset also have an influence on the total computation time. 

\begin{figure}[!tbp]
\centerline{\includegraphics[width=9cm]{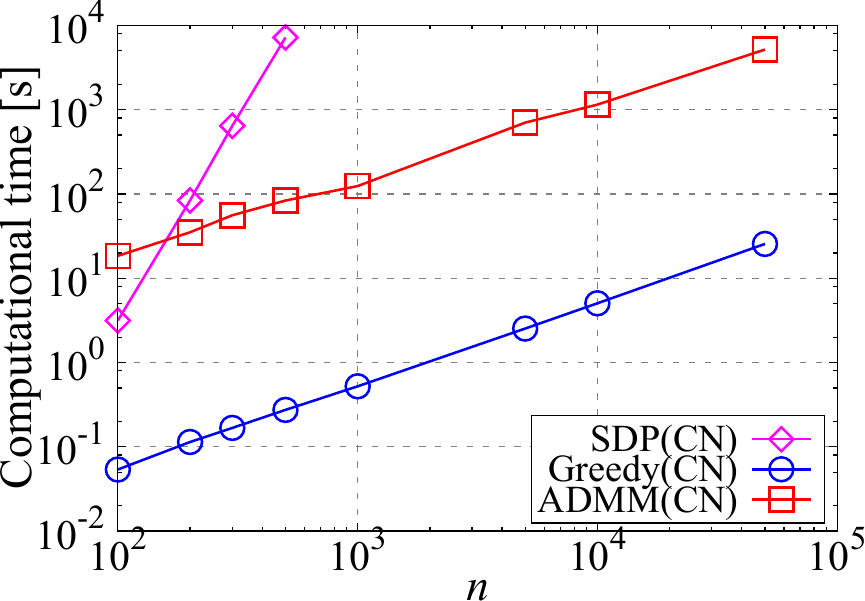}}
\caption{\reviewerA{Computational time with respect to the number of potential sensor locations for $p=30$.}}
\label{fig:time_n} 
\end{figure}

Fig.~\ref{fig:time_n} shows the comparison of computational time. The numerical test was carried out on a desktop personal computer (Intel$^{\textregistered}$ Xeon$^{\textregistered}$ W-2295 3.0~GHz CPU with 256~GB RAM). The size of dataset was ${\mathbf X}_{\mathrm{data}}\in\mathbb{R}^{n\times 50}$. The number of potential sensor locations $n$ was changed from $10^2$ to $5\times10^4$, and the average computational time and objective value of 100 times computations with different datasets were compared. The number of selected sensors was fixed at $p=30$. 

The computational time for all methods increases as $n$ increases. The computational time of Greedy(CN) is much less than that of SDP(CN) and ADMM(CN). Although the computational time of SDP(CN) is less than that of ADMM(CN) at $n=10^2$, it rapidly increases as $n$ increases. The computational complexity of SDP(CN) is roughly given by $\mathcal{O}\left(n^{4.5}\right)$ \cite{liu2016sensor,nemirovski2004interior}, if the number of latent variables is much less than that of potential sensor locations. Consequently, approximately $10^4$~s is required even if the potential sensor location was $n=500$ when SDP(CN) is used. In addition, approximately 200~GB of random access memory is required for the problem with 500 sensor candidates. On the other hand, an increase in the computational time with respect to $n$ of ADMM(CN) is less than that of SDP(CN) and is at the same level as that of Greedy(CN), and problems with $n=5\times10^4$ can be treated.

It should be noted that the 
present method assumes that the noise covariance matrix can be approximated by a low-rank form based on the framework proposed by Yamada et al.~\cite{yamada2021fast,yamada2022greedy}. Therefore, an additional computational cost might be 
required when using a more general noise model, e.g. an inverse-distance noise model in which the correlated measurement noise is inversely proportional to the distance between sensors.

\subsection{A-optimality Criterion and Reconstruction Error}
\begin{figure}[!tbp]
\centerline{\includegraphics[width=9cm]{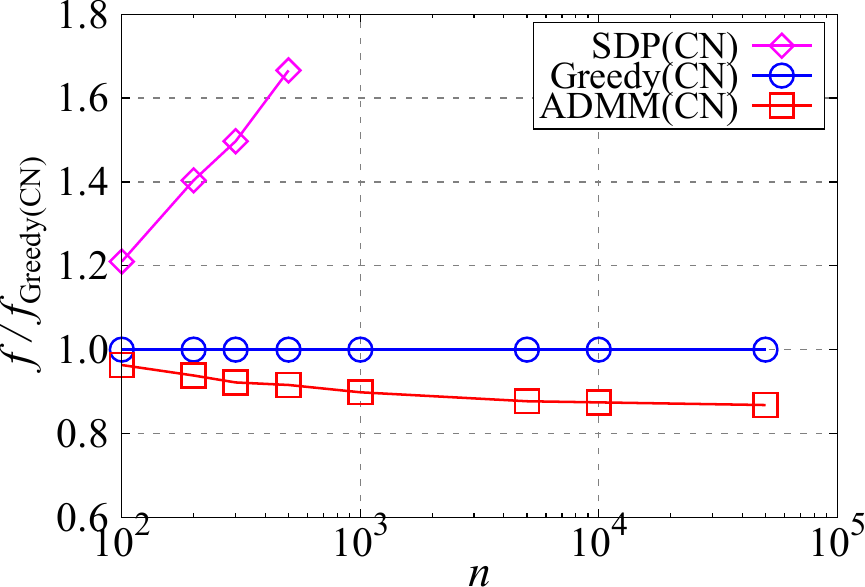}}
\caption{\reviewerA{Objective value with respect to the number of potential sensor locations for $p=30$.}}
\label{fig:objnorm_n} 
\end{figure}

\reviewerA{Fig.~\ref{fig:objnorm_n} shows the trace of the inverse of the FIM with respect to the number of potential sensor locations for $p=30$. The size of the dataset was ${\mathbf X}_{\mathrm{data}}\in\mathbb{R}^{10,000\times 100}$. The objective value was normalized by that obtained by Greedy(CN). Although SDP(CN) \cite{liu2016sensor} is quite general, the objective value obtained by SDP(CN) is larger, thus, worse than that obtained by Greedy(CN) and ADMM(CN). In addition, obtained objective value using SDP(CN) gets worse as the number of potential sensor locations increases. The computational time of ADMM(CN) is longer than that of Greedy(CN), as illustrated in Fig.~\ref{fig:objnorm_n}, but the objective value obtained is superior to that obtained by Greedy(CN) under the tested conditions.}

\begin{figure}[!tbp]
\centerline{\includegraphics[width=9cm]{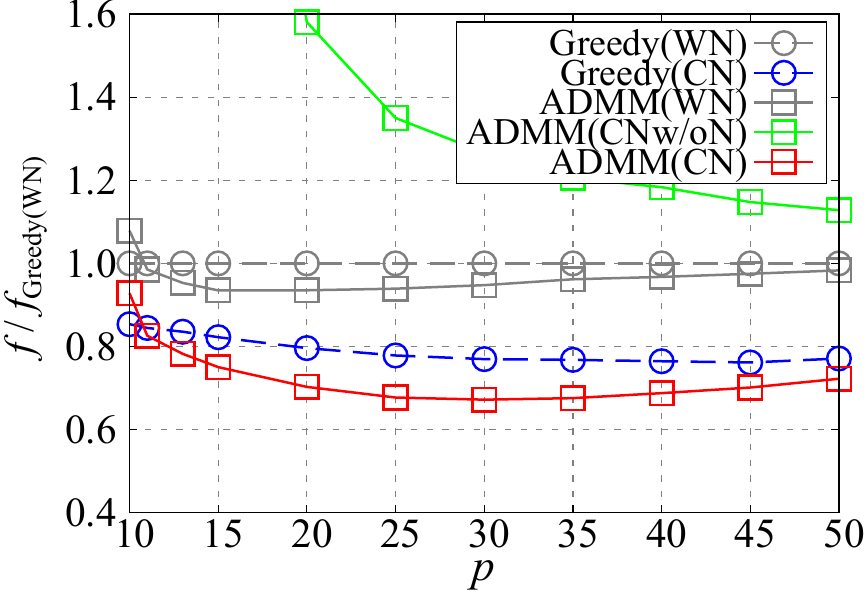}}
\caption{Comparison of objective values obtained in the experiment using the artificial dataset.}
\label{fig:trinv-rand} 
\end{figure}

Fig.~\ref{fig:trinv-rand} shows the trace of the inverse of the FIM with respect to the number of selected sensors. Here, the size of dataset was ${\mathbf X}_{\mathrm{data}}\in\mathbb{R}^{10,000\times 100}$. This figure illustrates that the objective values obtained by the methods that consider correlated measurement noise are smaller than those obtained by the methods that consider only white noise, which is an expected result.
The objective value obtained by ADMM(CN) is better than that obtained by Greedy(CN) and is the best among the compared methods, except for $p=10$. The objective values obtained by Greedy(WN) and ADMM(WN) are larger because the noise covariance matrix $\mathbf{R}$ is not considered in the sensor selection. The objective value obtained by ADMM(CNw/oN) is larger than that obtained by the method that does not consider correlated noise, even though the method considers correlated noise in the sensor selection.

As described in Section~\ref{sec:normalization}, the difference between ADMM(CN) and ADMM(CNw/oN) is described by the difference in their objective functions \eqref{eq:objGL0fnorm} and \eqref{eq:objorgGL0fnorm}. In the case of ADMM(CNw/oN), the noise covariance matrix is considered in the first term of the objective function, but it is not considered in the sparsity-promoting term. In the case of ADMM(CN), on the other hand, the sensor candidate matrix $\mathbf{U}$ is normalized by the noise weighting term $\mathbf{R}_{\rm d}$, which is the diagonal matrix with the diagonal entries of the noise covariance matrix $\mathbf{R}$. Therefore, the noise intensity is considered in the computation of the sparsity-promoting term for ADMM(CN). The consideration of off-diagonal entries is more difficult because the effect of correlated noise cannot be accurately evaluated unless the sensor locations are determined, as discussed in Section~\ref{sec:diff}. Therefore, it is difficult to consider the correlated measurement noise in the sparsity-prompting term, and the off-diagonal entries are not fully considered even in ADMM(CN). It should be noted that the noise considered in the present study has a certain size structure because the noise model consists of truncated modes. Hence, not only the uncorrelated noise but also part of the component of correlated noise will be considered by normalization using the noise weighting term, even if only the diagonal component of the noise covariance matrix is considered. The result of this numerical experiment indicates that the normalization of the sensor candidate matrix by the noise weighting term is effective and suggests that consideration of the intensity of noise in the thresholding process is important for optimization using proximal methods.

\begin{figure}[!tbp]
\centerline{\includegraphics[width=9cm]{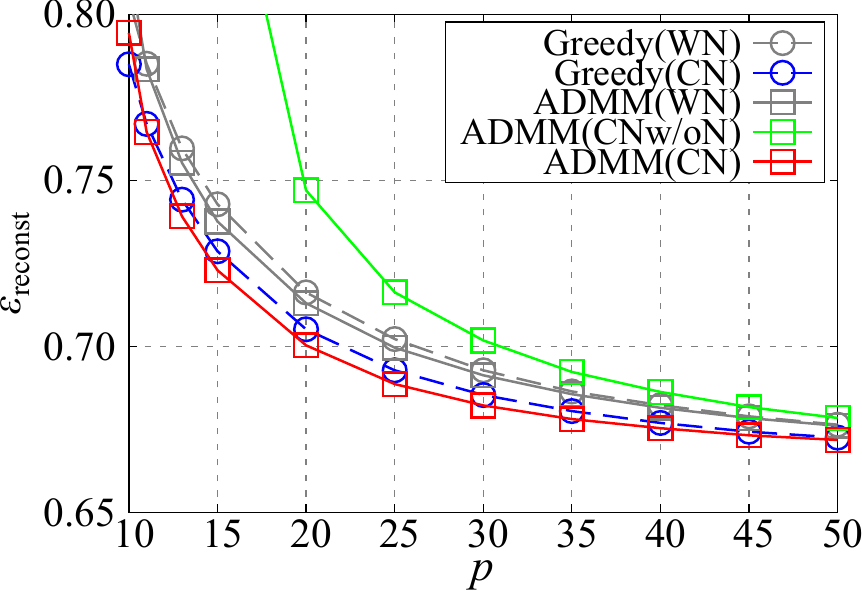}}
\caption{Reconstruction error for an artificial dataset.}
\label{fig:reconst-rand} 
\end{figure}

Fig.~\ref{fig:reconst-rand} shows the reconstruction error, which is the difference between the reconstructed and original data, defined as follows:
\begin{align}
    \epsilon_{\rm reconst}=\frac{||\mathbf{X}_\mathrm{data}-\mathbf{U\reviewerD{\tilde{Z}}}||_{\rm F}}{||\mathbf{X}_\mathrm{data}||_{\rm F}},
    \label{eq:recosterr}
\end{align}
\noindent
where $\tilde{\mathbf{Z}}$ is the estimated coefficients of modes obtained by \eqref{eq:WLS_estimation}. The notation $\left|\left|\,\cdot\,\right|\right|_{\rm F}$ indicates the Frobenius norm of a matrix. The reconstruction error is reduced by considering the correlated measurement noise, and the smallest reconstruction error is obtained for ADMM(CN). As pointed out with regard to Fig.~\ref{fig:trinv-rand}, the performance of ADMM(CNw/oN) is the worst, even though correlated noise is considered. The trends of the reconstruction error are the same as those of the objective value because it is an ideal problem setting.

The greedy algorithm can only provide a sensor subset consisting of $p$ sensors only by adding a new sensor that gives the largest increment of the objective to the sensor subset consisting of $p-1$ sensors. The influence of this constraint on greedy strategy is relatively small at $p<r$ due to the nature of the undersampling condition, but the constraint tends to degrade the performance of the sensor subset obtained under oversampling conditions. On the other hand, the ADMM-based method can obtain different suboptimal solutions for each number of sensors, similar to the convex relaxation method. Hence, the performance of the sensor subset obtained by ADMM(CN) is higher than that obtained by Greedy(CN) at larger $p$. However, the difference in the objective value and reconstruction error depending on the method decreases as the number of selected sensors $p$ increases. This is because as the number of sensors $p$ increases, the importance of location and combination of each selected sensor becomes smaller than when the number of selected sensors is small. \reviewerD{On the other hand, the condition on $p=10$ is on the constraint of the ADMM-based method, and thus, the optimization can be severe and the quality of the solution deteriorates.} Hence, the difference in the performance between ADMM(CN) and Greedy(CN) is the largest at around $p=25$.

\section{Application to Data-driven Sensor Selection}
The proposed method was applied to a data-driven sensor selection problem, and comparisons with the previously proposed method were conducted. The adopted dataset was the NOAA-OISST V2 dataset, which comprises weekly global sea surface temperature measurements between 1990 and 2000 \cite{noaa}. The dataset used consisted of 520 snapshots with 44,219 points of a spatial grid. The temperature data for 10 years were split into training data $\mathbf{X_{\rm train}}$ (80\% of the dataset) and the test data $\mathbf{X_{\rm test}}$ (20\% of the dataset), and five-fold cross-validation was conducted.

The data matrix $\mathbf{X}_\mathrm{data}\in\mathbb{R}^{n\times m}$, which consisted of $m$ snapshots with a spatial dimension of $n$, was decomposed into a left singular matrix $\mathbf{U}\in\mathbb{R}^{n\times m}$, which shows the spatial modes; a diagonal matrix of singular values $\mathbf{S}\in\mathbb{R}^{m\times m}$; and a right singular matrix $\mathbf{V}\in\mathbb{R}^{m\times m}$, which shows the temporal modes. Here, the spatial dimension and the number of snapshots for this dataset were $n=44,219$ and $m=416$, respectively. The dimensional reduction was conducted by the truncated SVD, and the rank-$r$ reduced-order modeling of a data matrix is given as the $r$-rank approximation $\mathbf{X}_\mathrm{data}\approx\mathbf{U}_{1:r_1}\mathbf{S}_{1:r_1}\mathbf{V}_{1:r_1}^{\mathsf T}$ \cite{eckart1936approximation}. The reduced-order model was constructed by retaining the leading-$r_1$ singular values and vectors, where $r_1=10$ in the present experiment. The noise covariance matrix was constructed using truncated modes up to $r_2=40$ modes based on \eqref{eq:ncovlow}. The objective value \eqref{eq:obj_tr} and the reconstruction error \eqref{eq:recosterr} were evaluated. The initial values of $\gamma_{\rm init}$ for ADMM(WN) and ADMM(CN) were set at 0.7 and 1.2, respectively, and the value was multiplied by $\eta=0.99$ every 5000 iterations, similar to the numerical experiments in Section~\ref{sec:rand}. \reviewerA{The latent variables $\mathbf{Z}_{1:r} (=\mathbf{S}_{1:r_1}\mathbf{V}_{1:r_1}^{\mathsf T})$ in this problem are mode coefficients of the spatial modes of the time variation in the temperature field $\mathbf{U}_{1:r}$. The mode coefficients $\mathbf{Z}_{1:r}$ indicate the time variation in the strength of each mode.}

\begin{figure}[!tbp]
\centerline{\includegraphics[width=9cm]{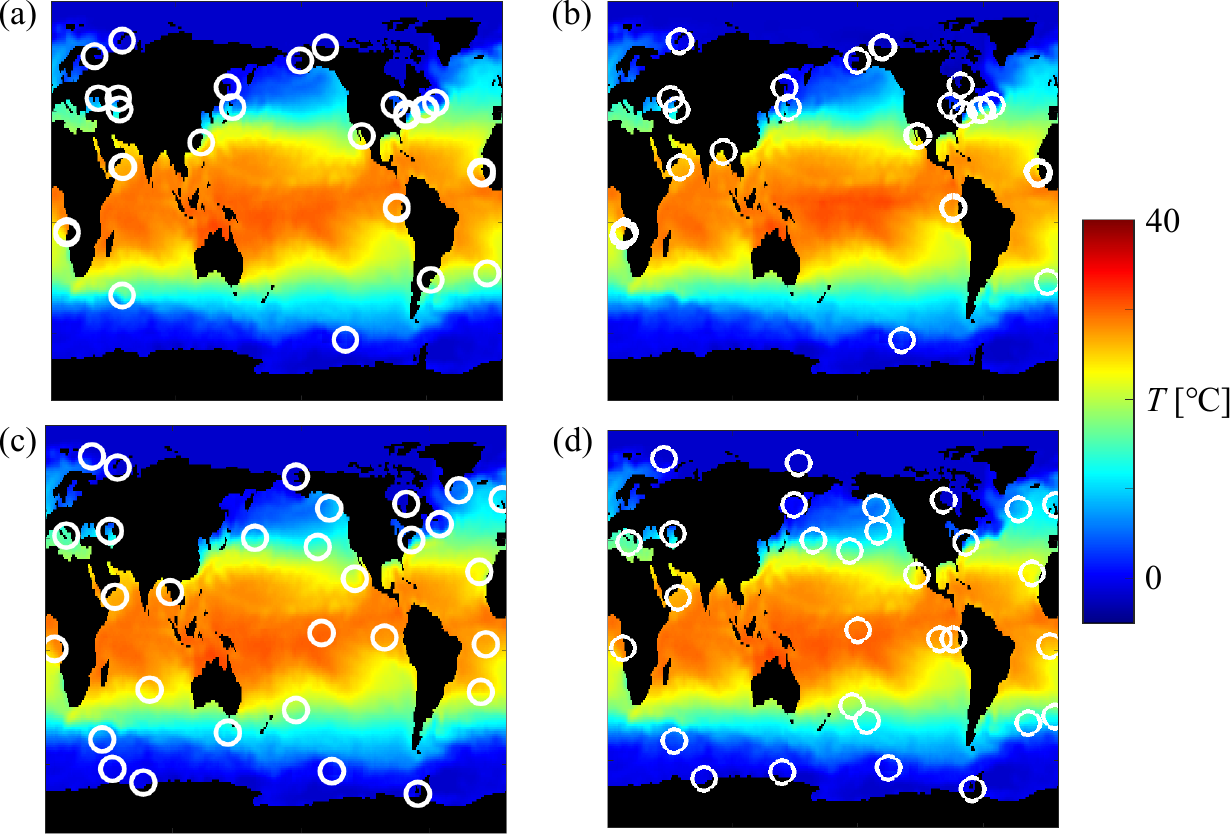}}
\caption{Locations of selected sensors at $p=30$. (a) Greedy(WN), (b) ADMM(WN), (c) Greedy(CN), (d) ADMM(CN). The white open circles represent the locations of activated sensor locations.}
\label{fig:sensor-noaa} 
\end{figure}

Fig.~\ref{fig:sensor-noaa} illustrates the locations of the 30 sensors selected by the greedy and ADMM-based methods. The white open circles represent sensor locations. There is a clear difference in the locations of the selected sensors obtained by the methods considering white noise and correlated noise. The sensors selected by the methods assuming white noise are aggregated into several regions. In contrast, the location of sensors is dispersed by considering the correlated measurement noise. Observations using the latter sensor subset can be more informative than observations with the former subset.

\begin{figure}[!tbp]
\centerline{\includegraphics[width=9cm]{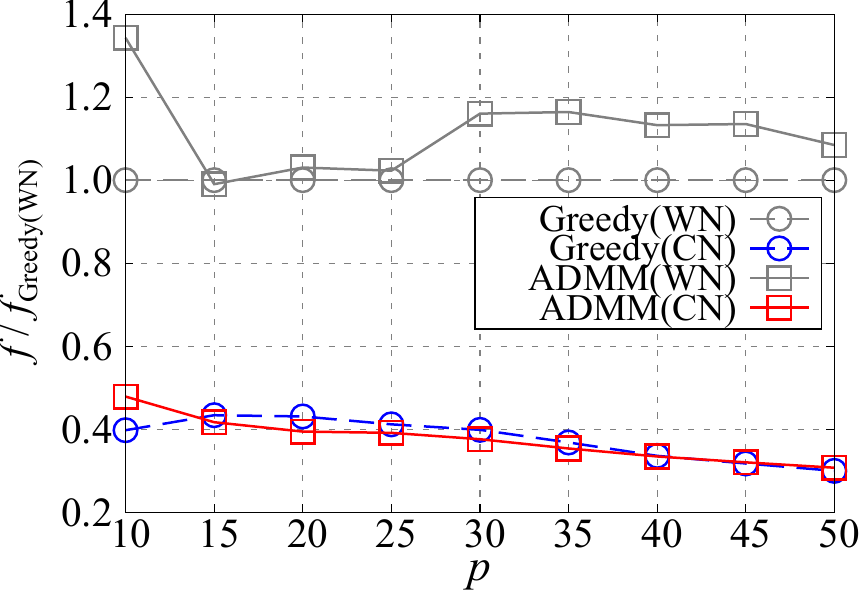}}
\caption{Comparison of objective values obtained in experiment using practical dataset.}
\label{fig:trinv-noaa} 
\end{figure}

The objective values are shown in Fig.~\ref{fig:trinv-noaa}. As discussed in Section~\ref{sec:rand}, the objective values obtained by Greedy(CN) and ADMM(CN) are smaller than those obtained by Greedy(WN) and ADMM(WN). 
The objective value obtained by ADMM(CN) is better than that obtained by Greedy(CN) at around $15\leq p\leq 40$ and is worse at smaller and larger values of $p$. For $p=10$, the condition is on the constraint of the ADMM-based method, and thus, the optimization can be severe and the quality of the solution deteriorates. A similar trend for the ADMM-based method has been reported in a previous study \cite{nagata2021data} and can also be seen in the convex relaxation method \cite{joshi2009sensor} (see \cite{saito2021determinantbased,nakai2021effect,nagata2021data}).

\begin{figure}[!tbp]
\centerline{\includegraphics[width=9cm]{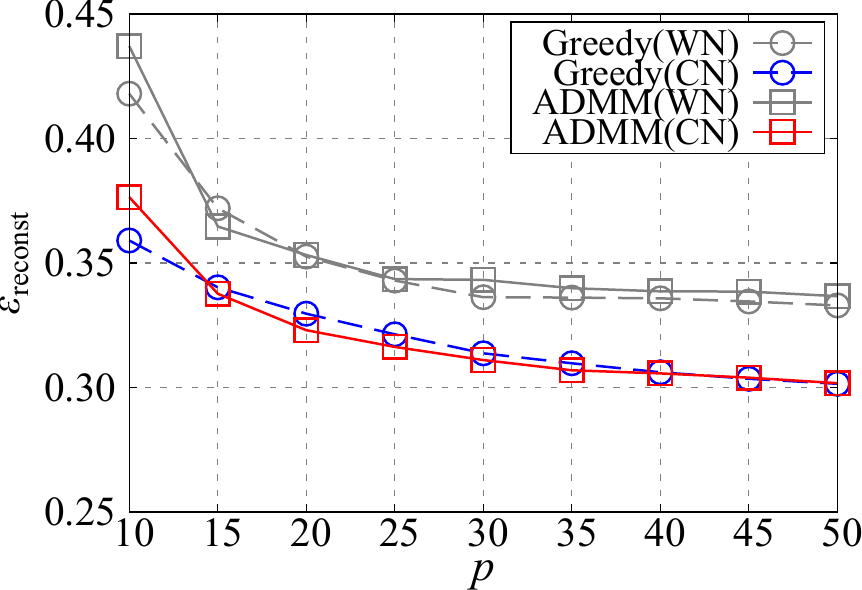}}
\caption{Reconstruction error for the NOAA-OISST dataset.}
\label{fig:reconst-noaa}
\end{figure}

Fig.~\ref{fig:reconst-noaa} shows that the reconstruction error is reduced by considering the correlated measurement noise, as was seen for the numerical experiment with the artificial dataset. ADMM (CN) achieved the smallest reconstruction error in the range $15\leq p\leq 40$. As discussed in the numerical experiment using the artificial dataset, the importance of the location and combination of each activated sensor gradually decreases as the number of sensors $p$ increases. Therefore, the difference in the reconstruction error between ADMM(CN) and Greedy(CN) decreases for larger $p$. In addition, the performance of the ADMM-based methods deteriorates at $p=10$ compared to greedy methods.

\section{Conclusions}
The present paper proposes a data-driven sensor selection method for high-dimensional nondynamical systems with correlated measurement noise. The proposed method is based on the proximal splitting algorithm and the A-optimal design of experiments. The sensor locations are determined by minimizing the trace of the inverse of the FIM with group $\ell_0$ pseudo-norm constraint. The proposed method can avoid the difficulty of sensor selection with strongly correlated measurement noise, namely, that the sensor locations must be known in advance for selecting the sensor locations, as described in Section~\ref{sec:diff}. The problem can be efficiently solved with ADMM, and the computational complexity is $\mathcal{O}\left(n\right)$ when the method is combined with a low-rank expression of the measurement noise model \cite{yamada2021fast,yamada2022greedy}. Therefore, the proposed method can treat large-scale problems that have more than 10,000 potential sensor locations, such as data-driven sensor selection problems.

The performance of the proposed method was compared with those of the previously proposed methods by the numerical experiments using artificial and practical datasets. The adopted practical dataset was NOAA-OISST V2, which contains weekly mean sea surface temperatures.

The results of numerical experiments using the artificial dataset and NOAA-OISST dataset showed that the performance of ADMM(CN) is the best among the compared methods in terms of the objective value and reconstruction error. The ADMM-based method without normalization by the noise weighting term [ADMM(CNw/oN)] was also tested to investigate the effectiveness of noise normalization in the thresholding process. Although ADMM(CNw/oN) considers correlated noise, the obtained objective value was larger than those obtained by methods that do not consider correlated noise. This indicates that the normalization of the sensor candidate matrix by the noise weighting term is effective and suggests that consideration of the noise intensity in the thresholding process is essential for optimization using proximal methods.


\reviewerA{The previous methods considering the correlated measurement noise require the calculation of the precision matrix (inverse of the noise covariance matrix) which should be constructed by the noise information over only selected sensors. In particularly, the method based on convex relaxation requires much larger calculation cost or a complicated formulation with limitations. There was no published method to formulate the continuous optimization problem for sensor selection that can be solved by a method with a computational complexity less than a cubic order of the number of potential sensor locations. Hence, it was difficult to employ the previously proposed methods for high-dimensional data including such correlated measurement noise.} Although the computational cost becomes quite large when using a more general noise model, the proposed method can achieve low-cost nonconvex sensor optimization considering correlated measurement noise by means of a low-rank expression of correlated measurement noise \cite{yamada2021fast}. Further extension of the complex objective function and its constraints is expected using the framework proposed in the present study.

\section*{Acknowledgement}
This work was supported by JST CREST Grant Number JPMJCR1763, JST FOREST Grant Number JPMJFR202C, JST PRESTO Grant Number JPMJPR21C4, and JSPS KAKENHI Grant Number JP22H03610.

\bibliographystyle{IEEEtran}
\bibliography{main.bib}

\begin{IEEEbiography}
[{\includegraphics[width=1in,height=1.25in,clip,keepaspectratio]{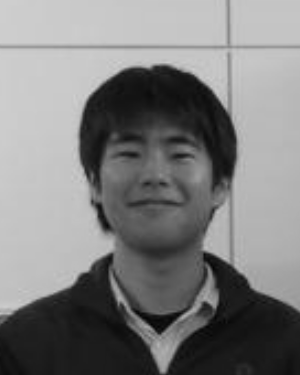}}]{Takayuki Nagata} received the B.S. and M.S. degrees in mechanical and aerospace engineering from Tokai University, Hiratsuka, Japan, in 2015 and 2017, respectively. He received the Ph.D. degree in aerospace engineering from Tohoku University, Sendai, Japan, in 2020. From 2018 to 2020, he was a Research Fellow of the Japan Society for the Promotion of Science (JSPS) at Tohoku University, Japan. He is currently a Project Assistant Professor at Tohoku University, Sendai, Japan. 
\end{IEEEbiography}

\begin{IEEEbiography}
[{\includegraphics[width=1in,height=1.25in,clip,keepaspectratio]{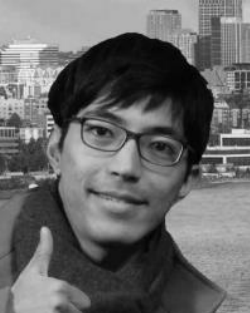}}]{Keigo Yamada} received the B.S. degree in physics and M.S. degree in aerospace engineering from Tohoku University, Sendai, Japan, in 2019. He is currently a Ph.D. student with the Department of Aerospace Engineering with Tohoku University, Sendai, Japan.
\end{IEEEbiography}

\begin{IEEEbiography}
[{\includegraphics[width=1in,height=1.25in,clip,keepaspectratio]{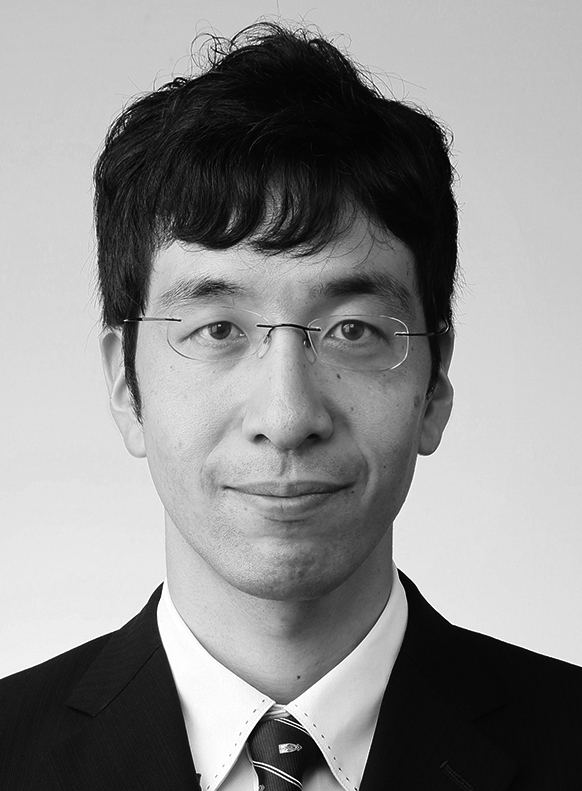}}]{Taku Nonomura} received the B.S. degree in mechanical and aerospace engineering from Nagoya University, Nagoya, Japan, in 2003, and the Ph.D. degree in aerospace engineering from the University of Tokyo, Japan in 2008. He is currently an Associate Professor with the Department of Aerospace Engineering at Tohoku University, Sendai, Japan.
\end{IEEEbiography}

\begin{IEEEbiography}
[{\includegraphics[width=1in,height=1.25in,clip,keepaspectratio]{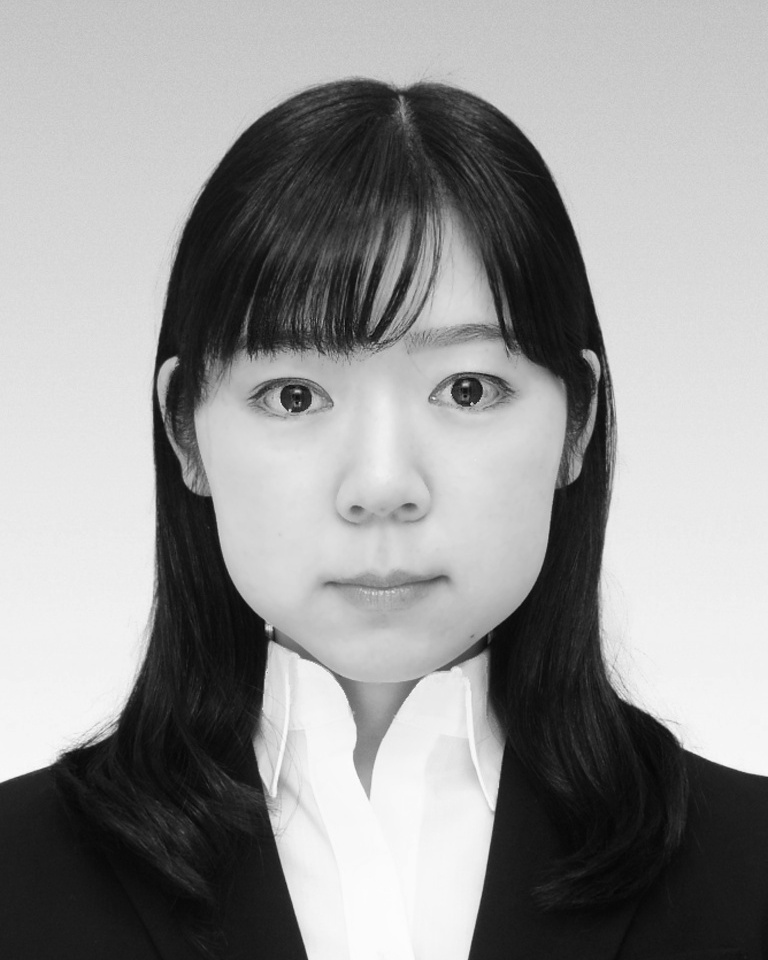}}]{Kumi Nakai} received the Ph.D. degree in mechanical systems engineering from Tokyo University of Agriculture and Technology, Tokyo, Japan, in 2020. From 2017 to 2020, she was a Research Fellow of the Japan Society for the Promotion of Science (JSPS) at Tokyo University of Agriculture and Technology, Tokyo, Japan. She is currently a researcher with National Institute of Advanced Industrial Science and Technology, Japan.
\end{IEEEbiography}

\begin{IEEEbiography}
[{\includegraphics[width=1in,height=1.25in,clip,keepaspectratio]{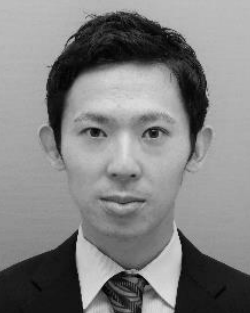}}]{Yuji Saito} received the B.S. degree in mechanical engineering, and the Ph.D. degree in mechanical space engineering from Hokkaido University, Japan, in 2018. He is currently an Assistant Professor with the Department of Aerospace Engineering at Tohoku University, Sendai, Japan.
\end{IEEEbiography}

\begin{IEEEbiography}
[{\includegraphics[width=1in,height=1.25in,clip,keepaspectratio]{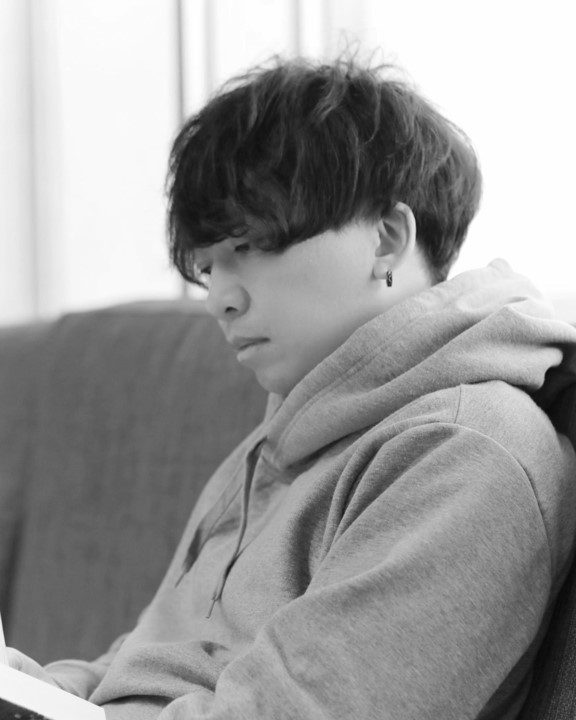}}]{Shunsuke Ono} (S’11–M’15) received a B.E. degree in Computer Science in 2010 and M.E. and Ph.D. degrees in Communications and Computer Engineering in 2012 and 2014 from the Tokyo Institute of Technology, respectively.

From April 2012 to September 2014, he was a Research Fellow (DC1) of the Japan Society for the Promotion of Science (JSPS). He is currently an Associate Professor in the Department of Computer Science, School of Computing, Tokyo Institute of Technology. From October 2016 to March 2020 and from October 2021 to present, he was/is a Researcher of Precursory Research for Embryonic Science and Technology (PRESTO), Japan Science and Technology Corporation (JST), Tokyo, Japan. His research interests include signal processing, image analysis, remote sensing, mathematical optimization, and data science.

Dr. Ono received the Young Researchers’ Award and the Excellent Paper Award from the IEICE in 2013 and 2014, respectively, the Outstanding Student Journal Paper Award and the Young Author Best Paper Award from the IEEE SPS Japan Chapter in 2014 and 2020, respectively, the Funai Research Award from the Funai Foundation in 2017, the Ando Incentive Prize from the Foundation of Ando Laboratory in 2021, and the Young Scientists’ Award from MEXT in 2022. He has been an Associate Editor of IEEE TRANSACTIONS ON SIGNAL AND INFORMATION PROCESSING OVER NETWORKS since 2019.
\end{IEEEbiography}

\end{document}